\documentclass[%
 reprint,
superscriptaddress,
 amsmath,amssymb,
 aps,
 prd,
]{revtex4-2}

\usepackage{graphicx}
\usepackage{dcolumn}
\usepackage{bm}
\RequirePackage{graphicx}
\RequirePackage{subfigure}
\usepackage{rotating}
\usepackage{amsmath}
\usepackage{mathtools}
\RequirePackage[colorlinks,citecolor=blue,urlcolor=blue,linkcolor=blue]{hyperref}
\usepackage[]{units}
\usepackage{xspace}
\usepackage[switch]{lineno}
\usepackage{comment}

\usepackage[switch]{lineno} 
\usepackage{lipsum}

\usepackage{tabularx} 

\begin{document}

\preprint{APS/123-QED}

\title{Exploring the keV-scale physics potential of CUORE}
\author{D.~Q.~Adams}
\affiliation{Department of Physics and Astronomy, University of South Carolina, Columbia, SC 29208, USA}
\author{C.~Alduino}
\affiliation{Department of Physics and Astronomy, University of South Carolina, Columbia, SC 29208, USA}
\author{K.~Alfonso}
\affiliation{Center for Neutrino Physics, Virginia Polytechnic Institute and State University, Blacksburg, Virginia 24061,USA}
\author{A.~~Armatol}
\affiliation{Nuclear Science Division, Lawrence Berkeley National Laboratory, Berkeley, CA 94720, USA}
\author{F.~T.~Avignone~III}
\affiliation{Department of Physics and Astronomy, University of South Carolina, Columbia, SC 29208, USA}
\author{O.~Azzolini}
\affiliation{INFN -- Laboratori Nazionali di Legnaro, Legnaro (Padova) I-35020, Italy}
\author{G.~Bari}
\affiliation{INFN -- Sezione di Bologna, Bologna I-40127, Italy}
\author{F.~Bellini}
\affiliation{Dipartimento di Fisica, Sapienza Universit\`{a} di Roma, Roma I-00185, Italy}
\affiliation{INFN -- Sezione di Roma, Roma I-00185, Italy}
\author{G.~Benato}
\affiliation{Gran Sasso Science Institute, L'Aquila I-67100, Italy}
\affiliation{INFN -- Laboratori Nazionali del Gran Sasso, Assergi (L'Aquila) I-67100, Italy}
\author{M.~Beretta}
\affiliation{Dipartimento di Fisica, Universit\`{a} di Milano-Bicocca, Milano I-20126, Italy}
\affiliation{INFN -- Sezione di Milano, Milano I-20133, Italy}
% ...

\collaboration{CUORE Collaboration}

\date{\today}

\begin{abstract}
We present the analysis techniques developed to explore the keV-scale energy region of the CUORE experiment, based on more than 2 tonne yr of data collected over 5 years. By prioritizing a stricter selection over a larger exposure, we are able to optimize data selection for thresholds at 10 keV and 3 keV with 691 kg yr and 11 kg yr of data, respectively.
We study how the performance varies among the 988-detector array with different detector characteristics and data taking conditions.
We achieve an average baseline resolution of 2.54 $\pm$ 0.14 keV FWHM and 1.18 $\pm$ 0.02 keV FWHM for the data selection at 10 keV and 3 keV, respectively. The analysis methods employed reduce the overall background by about an order of magnitude, reaching 2.06 $\pm$ 0.05 counts/(keV kg days) and 16 $\pm$ 2 counts/(keV kg days) at the thresholds of 10 keV and 3 keV.
We evaluate for the first time the near-threshold reconstruction efficiencies of the CUORE experiment, and find these to be 50 $\pm$ 2 \% and 26 $\pm$ 4 \% at 10 keV and 3 keV, respectively.
This analysis provides crucial insights into rare decay studies, new physics searches, and keV-scale background modeling with CUORE.
We demonstrate that tonne-scale cryogenic calorimeters can operate across a wide energy range, from keV to MeV, establishing their scalability as versatile detectors for rare event and dark matter physics. These findings also inform the optimization of future large mass cryogenic calorimeters to enhance the sensitivity to low-energy phenomena.
\end{abstract}

\maketitle


\section{Introduction}
\label{intro}
The capability of building tonne-scale underground experiments is crucial to investigate new physics scenarios in the fields of neutrinos and dark matter.
The development of large mass cryogenic calorimeters has reached maturity with the Cryogenic Underground Observatory for Rare Events (CUORE), which is the first experiment to demonstrate the feasibility of operating this particle detection technology at the tonne scale and maintaining stable data taking conditions over several years \cite{nature}. 
The main physics goal of CUORE is to search for neutrinoless double beta decay (0$\nu\beta\beta$) of $\mathrm{^{130}Te}$. This process would manifest as a monochromatic peak at the Q-value of the nuclear reaction, $\sim$2.5 MeV for $\mathrm{^{130}Te}$ \cite{0nureview}. Its discovery would assess the neutrino as a Majorana particle, providing insights into the matter-antimatter asymmetry of the Universe enigma \cite{baryonasym}.

To perform this search, CUORE uses 988 natural TeO$_2$ cryogenic calorimeters operated at $\sim$10 mK. The low heat capacity allows the conversion of energy deposits to small increases in temperature. The seconds-long thermalization process is converted into an electrical pulse through thermistors, i.e. cryogenic sensors whose resistance varies with the temperature. This technology provides excellent resolution capabilities together with a wide operating energy range. CUORE has collected more than 2.5 tonne yr of TeO$_2$ exposure, 2 of which have been processed and analyzed, reaching an average energy resolution of $\sim$7.3 keV and a background level of 1.4 $\cdot$ 10$^{-2}$ counts/(keV kg year) at the $\mathrm{^{130}Te}$ 0$\nu\beta\beta$ Q-value \cite{cuorecollaboration2024nuhuntingseedmatterantimatter}.

Following this multi-year data taking campaign, we can leverage the experiment's low background and high exposure with its sensitivity to a broad energy range to perform searches for exotic physics in the keV regime. Among these, the dark matter puzzle is a long-standing mystery still unaddressed \cite{reviewdm}. The hunt for dark matter particle candidates demands high exposure and low background detectors sensitive to keV-scale energy deposits hypothesized by several new physics models. The CUORE experiment can provide insights into this challenge by exploiting the sharp energy resolution typical of cryogenic calorimeters ($\sim$keV at the baseline) and the years-long livetime of the data collection \cite{Li_2015,Li_2016,inelasticdm,LowEnergyTechniques}. A good energy resolution is beneficial in the search for axions \cite{axionreview}, which are predicted to release monochromatic energy deposits, and the long live-time allows us to search for annual rate modulation as predicted by WIMP dark matter models \cite{dmdirect,Freese_2013}. 

\begin{figure}[!htbp]
    \centering
    \includegraphics[scale=0.45]{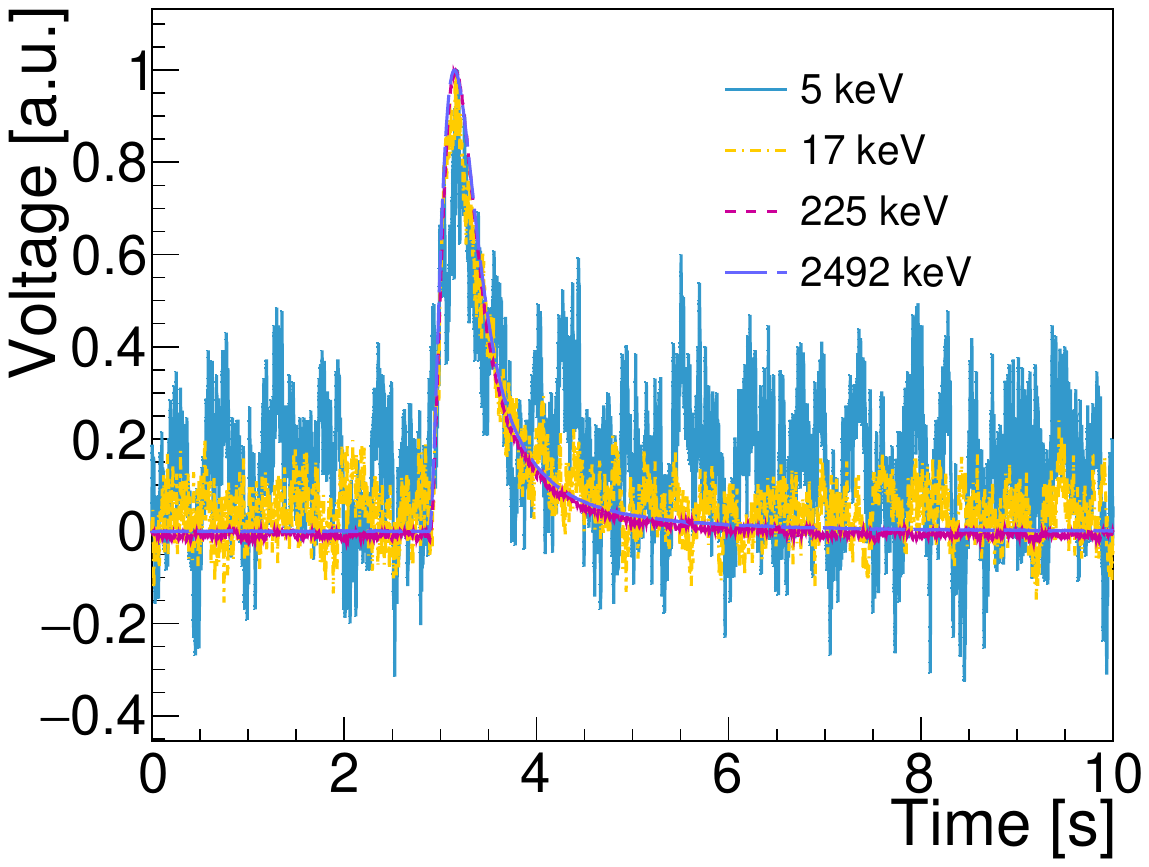}
    \caption{A sample of pulses from the same detector and operating conditions with different energies, decreasing from $\sim$2.5 MeV, in the energy region of interest for the 0$\nu\beta\beta$, to $\sim$5 keV, near CUORE threshold. The amplitude and the offset are normalized to be the same for all the pulses.}
    \label{pulses}
\end{figure}

Since standard analysis for 0$\nu\beta\beta$ search with CUORE relies on a 40 keV energy threshold, an optimization of the data processing is necessary to make use of the keV-scale data. The threshold of CUORE detectors is of the order of a few keV, so the spectrum we are interested in is polluted by noise and spurious events unless additional event selection criteria are applied. To explore this energy region, we developed dedicated analysis methods devoted to the selection of the highest quality data close to the trigger thresholds. 
As an example, we show in Fig. \ref{pulses} four events at different energies. These pulses demonstrate how the noise contribution becomes more prominent at lower energies. This affects several steps of the analysis, primarily the estimation of the pulse shape parameters needed to characterize an event and the detector's efficiencies. Moreover, this region of the energy spectrum is highly susceptible to spurious pulses, which originate from external sources, e.g. electronics and vibrations, inducing phonon excitation in the crystals.

In the following, we describe the analysis techniques applied to the 2 tonne yr data release of CUORE to explore keV-scale energy deposits. 
The overall impact of these techniques is the ability to reconstruct low energy deposits as a result of an improved resolution and a lower background, with the consequence of lower exposure and efficiency. We choose two classes of data selection: a more conservative one with a 10 keV threshold, and a stricter one at 3 keV.
This work aims to demonstrate the possibility of operating and studying a tonne-scale cryogenic calorimeter experiment such as CUORE over a wide energy range, spanning 3 orders of magnitude from a few keV, as presented in this work, to about 10 MeV \cite{nature,cuorecollaboration2024nuhuntingseedmatterantimatter,excitedstates,128te,120te,datadriven,fcp}.  

This paper is organized as follows. Sec. \ref{cuore} briefly describes the CUORE experiment and its data taking. Sec. \ref{da} describes all the analysis strategies and techniques that we developed specifically for low energy deposits. In particular, Sec. \ref{cal} details the verifications we performed about the energy reconstruction down to 3 keV, Sec. \ref{OT} introduces the offline triggering procedure applied to data, Sec. \ref{coincidences} highlights the improvements on tagging multi-sites events at the keV-scale, and Sec \ref{spurious} focuses on the rejection of spurious events polluting the near-threshold spectrum. Sec. \ref{dr} reports the results of the developed techniques in terms of selected exposure and energy resolution. Here we also investigates patterns in detector performance in the CUORE array over the course of data taking.  Sec. \ref{eff} reports the estimate of the energy-dependent efficiency due to event reconstruction and selection. Finally, Sec. \ref{spectra} shows the energy spectrum resulting from this study and highlights its main features.

\section{The CUORE Experiment}
\label{cuore}
CUORE,  located in the Laboratori Nazionali del Gran Sasso (LNGS) in Assergi, Italy, is composed of 988 5 cm $\times$ 5 cm $\times$ 5 cm  $\mathrm{natural}$ $\mathrm{TeO_2}$ crystal cryogenic calorimeters.
The crystals provide 742 kg of active mass and they are arranged in an array of 19 towers with 13 floors each mounted by copper holders secured by polytetraflouroethylene (PTFE). Towers are placed such that 12 of them form an outer layer containing the remaining 7 ones \cite{nature}. 

Each crystal is equipped with a Neutron Transmutation Doped Ge (NTD) sensor to detect thermal phonons generated by energy deposits in the crystal. In addition, each crystal has a Si chip used for injecting purely thermal signals of known voltage amplitude into the detector for thermal gain correction of long-term temperature drifts. 

The electrical connection between the crystal's cryogenic sensors and the front-end electronics is provided by copper traces deposited on flexible polyethylene naphthalate (PEN) substrates spanning the height of each tower. 

The experiment is operated in a custom $^{3}$He/$^{4}$He dilution refrigerator that allows the operation of the detectors at a millikelvin-scale temperature. It is a cryogen-free cryostat operated by means of 4 pulse tube cryocoolers. The CUORE cryostat comprises six nested copper vessels which thermalize at decreasing temperatures from room temperature down to 12-15 mK \cite{CUOREInfrastructure}. 

The detector array is suspended from a copper support plate placed at the center of the dilution refrigerator and anchored to a steel support through three mechanical insulators. These employ the Minus-K technology \cite{minusk} to tune the elastic constants 
of the suspension system so that the whole setup can tolerate heavy loads and act as a soft spring. The purpose of this system is to cut off the vibrations transferred to the cryostat. 

The experiment is shielded from cosmic rays with $\sim$3600 m.w.e. overburden from the Gran Sasso mountains \cite{nature}. Additionally, archaeological lead shields are mounted to the cryostat flanges to protect the experiment from natural radioactivity. Outside the cryostat, a room temperature lead shield and a neutron shield (made of polyethylene and a layer of boric acid) provide additional shielding from the side and from below.

To mitigate the experiment's background, radio-pure materials have been carefully chosen through specialized assay campaigns \cite{Alduino_2017}. Ultra-cleaning procedures have been designed and applied to these materials \cite{ALESSANDRIA201313} to address background arising from residual $\alpha$ decays on critical surfaces. Additionally, strict storage and handling protocols have been adopted to prevent recontamination during the assembly, installation, and commissioning of the detector array \cite{BUCCHERI2014130,Benato_2018}.
The data collection of the experiment started in 2017, and has currently collected more than 2.5 tonne yr of data.
CUORE data collection is divided into datasets which last about two months each and which are composed of day-long data-taking periods called runs.

The data acquisition system amplifies, filters, digitizes, and stores thermistor signals, allowing continuous data recording and offline triggering \cite{cuoredaq}. Its efficient processing enables multiple online triggers to identify signals, sample noise, and flag heater pulses.

The voltage measured by detectors is calibrated into energy by means of high-intensity $\gamma$ ray lines produced by $^{232}$Th - $^{60}$Co radioactive sources, which are deployed around the cryostat at the beginning and at the end of each dataset. 

The CUORE data taking can be divided into three configurations for the purposes of the low energy analysis. At the beginning CUORE was operated at $\sim$12 mK. After 2 years, the vibration isolators made by the Minus-K Technology were changed to a different configuration to improve the oscillations damping. Finally, during the last 2 years, CUORE was operated at a higher temperature of 15 mK \cite{nature,cuorecollaboration2024nuhuntingseedmatterantimatter}.

\section{Data Analysis}
\label{da}
Most of the data processing adopted for this study relies on the standard CUORE analysis chain, which is described in Ref. \cite{analysistechniques}. Before being triggered, a waveform's noise is de-correlated \cite{DenoisingInternalNote} from auxiliary devices (microphones, low frequency accelerometers and seismometers) placed in the cryostat hut, in order to mitigate the environmental contribution to vibrational noise \cite{DenoisingInternalNote}. Once triggered, waveforms are processed with the Optimum Filter (OF) \cite{Gatti:1986cw,Radeka:1966}, a matched filter used to suppress the frequencies most affected by the noise. This algorithm employs a detector response based on the average signal pulse shape and the average detector noise power spectrum extracted from data to build the transfer function on a detector basis. The amplitude estimate is corrected by the thermal gain to account for temperature instabilities. Finally, detectors are energy-calibrated by using dedicated runs with $^{232}$Th - $^{60}$Co source.

In order to analyze near-threshold events, we optimize the evaluation of some variables to account for the lower signal-to-noise ratio. For instance, the parameters of the algorithm to identify pile-up events by counting the number of pulses in an event window have been optimized to keV-scales and included in this analysis. However, the major difference with respect to other CUORE MeV-scale studies consists in the data selection, both in terms of single events and detector-dataset pairs.

\subsection{Validation data and energy scale calibration}
\label{cal}

An essential component for building and testing new analysis tools is having a sample of well-understood physics events or simulations. At higher energies, many ubiquitous $\gamma$-ray lines provide easily identifiable physics events. The possibility of exploiting these type of events at low energies is limited. However, low-energy X-rays from Tellurium are visible in the CUORE calibration spectra between $\sim$27 keV and $\sim$31 keV as single-site, i.e. depositing energy in a single crystal, labeled as M1 (Multiplicity 1) or double-site events (M2). The latter signature is more common in calibration data due to incoming radiation from external $^{232}$Th - $^{60}$Co source, which interacts primarily with Te atoms of the crystal's surfaces. The subsequent de-excitation of Te atoms produces X-rays that can travel to a neighboring crystal and be observed as an M2 event, making this the most prominent signature. These events were also observed in CUORE-0 and used to verify energy calibrations at low energy \cite{ThesisGPiperno,LowEnergyTechniques}. 

\begin{figure}[!htbp]   
\centering
    \includegraphics[scale=.4]{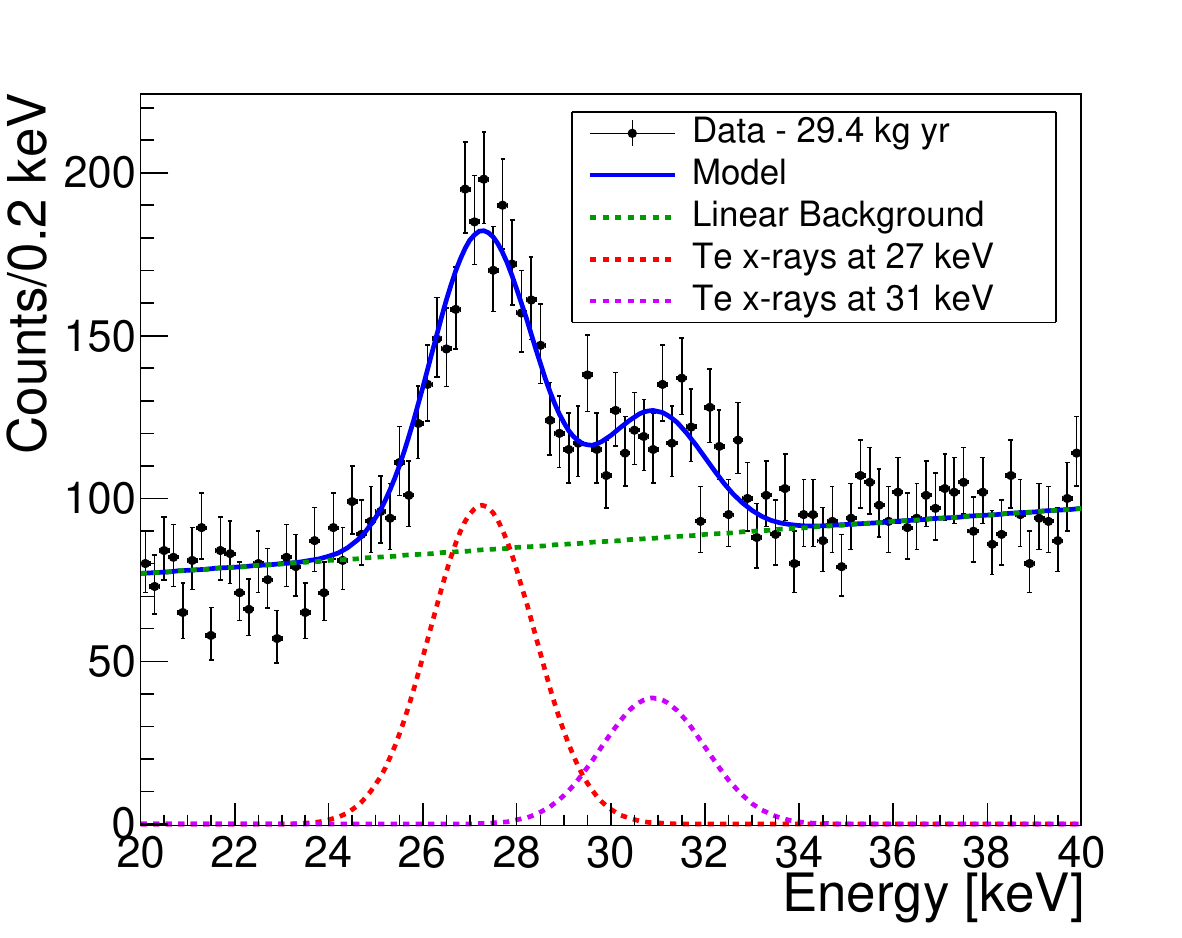}
    \caption{Data and model of Te X-rays peaks in a 20 keV width window. The events here reported belong to a sample calibration dataset of CUORE and they are selected to have multiplicity equal to 2 in order to enhance the signal-to-background ratio. The model consists of two Gaussian functions and a linear background. The fit to data resulted into a $\chi^2/\nu$ = 101/93, where $\nu$ is the number of degrees of freedom. The result in all of the other datasets is consistent with the example reported here.}
    \label{fig:TeXrays}
\end{figure}

CUORE's energy resolution does not allow for distinguishing between all of the eight most intense Te X-ray lines \cite{nucleardata,LowEnergyTechniques}. So, they are modeled as two Gaussian distributions with the same width but different mean value and with a floating amplitude ratio to account for an energy dependent attenuation length, as shown in Fig. \ref{fig:TeXrays}. Te X-rays are used as a source of reliable events while developing and validating multiple data analysis methods in this paper.

The energy calibration at low energies is extrapolated from the fit to higher energy $\gamma$ peaks. Specifically, a second order polynomial is fit to the amplitudes of the main peaks produced during the $^{232}$Th - $^{60}$Co sources deployment. We evaluate the quality of this calibration by fitting Te X-rays in the M2 calibration spectra for each dataset. We compute the energy shift with respect to the Te X-ray main peak, whose nominal value is taken as the average of the X-ray lines, weighted by their intensity. The average shift among the datasets results to be not significant within detector resolution, i.e. +0.05 $\pm$ 0.02 keV and +0.14 $\pm$ 0.06 keV for the detectors selected down to 10 and 3 keV, respectively. 

\begin{figure}[!htbp]   
	\centering
	\includegraphics[scale=.45]{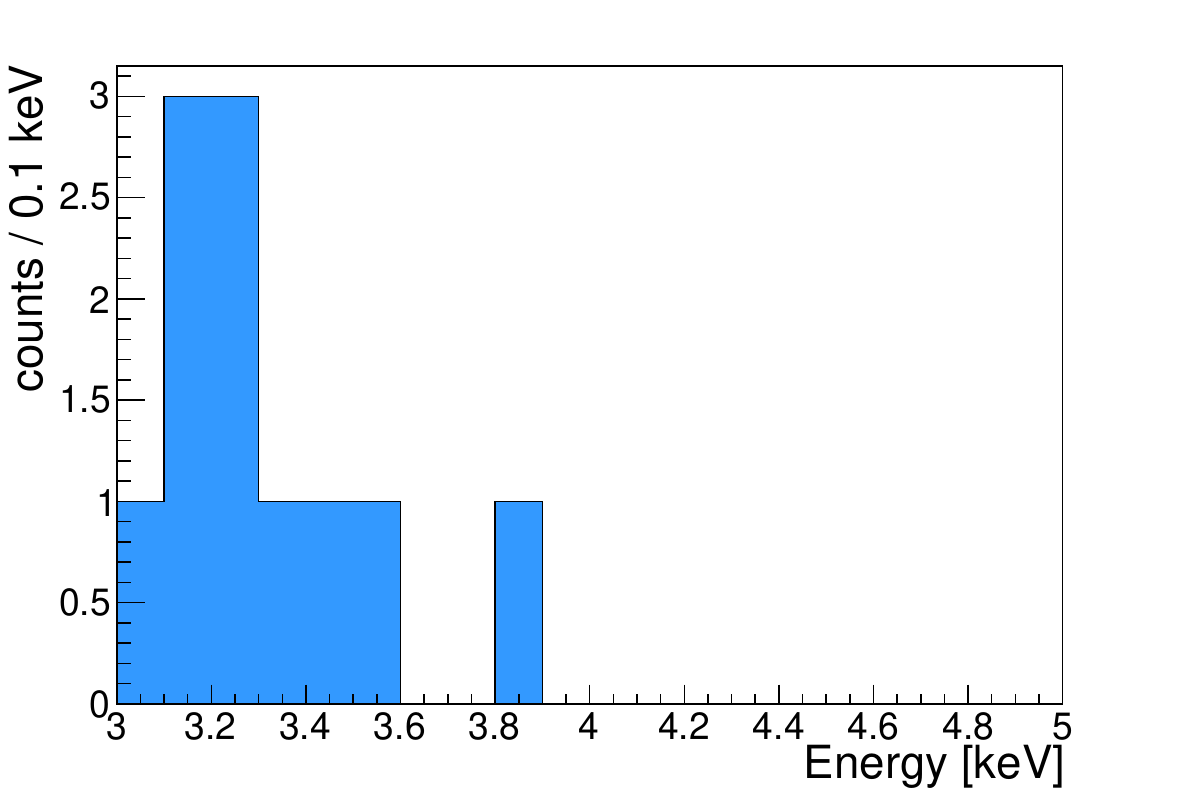}
	\caption[Calibration Check with 40K at low energy]{The histogram shows the distribution of low energy events coincident with the 1460 keV line from $^{40}$K line. The peak of the distribution coincide with the expected energy of the X-ray from the $^{40}$K electron capture. The total energy range of search extends up to 40 keV, and presents two more events not shown here for illustrative purposes. }
	\label{fig:calcheck}
\end{figure}

In order to validate the calibration at even lower energies, we looked for M2 events in coincidence with the $^{40}$K line at 1461 keV. Despite considering this line as purely M1 in most higher-energy CUORE analyses, there exists an associated X-ray emission at $\sim$3.2 keV \cite{lowEspectrum}. Among the detectors with trigger threshold lower than 3 keV, we identified ten events peaking at 3.2 keV, as shown in Fig. \ref{fig:calcheck}, confirming the reliability of the calibration near detector thresholds. As a comparison, given the time window of $\pm$15 ms and a rate of $\sim$2 mHz in both energy regions, the estimated number of random coincidences among the selected detectors along the used datasets is less than 4.

\subsection{Optimum Trigger}
\label{OT}
The first CUORE data processing step is to apply an offline trigger, the Optimum Trigger (OT) \cite{SDiDomizio_2011OT,LowEnergyTechniques}, to the continuous data-stream to identify events with an improved energy threshold. In particular, it triggers the waveforms upon optimizing the signal-to-noise ratio through digital filtering with the OF. The filter transfer function is built for each CUORE detector from derivative triggered data and applied on its continuous data stream; subsequently events are triggered. A signal is flagged by the OT if the filtered amplitude exceeds the filtered noise resolution by a factor $n$. The parameter $n$ is configurable on a detector basis, while the filtered noise resolution is known a priori from the filter transfer function of the given detector \cite{BrancaContour_2020}.

We evaluate the efficiency by applying the trigger to artificial pulses of various energies, based on a signal template, injected into noise waveforms from background data. The noise is randomly sampled every minute and required to have no pulses. The efficiency as a function of energy $\varepsilon(E)$ is modeled using the Gaussian cumulative density function to determine the trigger threshold, $E_{thr}$, set where $\varepsilon(E_{thr}) = 90\%$. We set the trigger threshold on each detector for every dataset.

\begin{figure}[!htbp]
	\centering
	\includegraphics[width=0.45\textwidth]{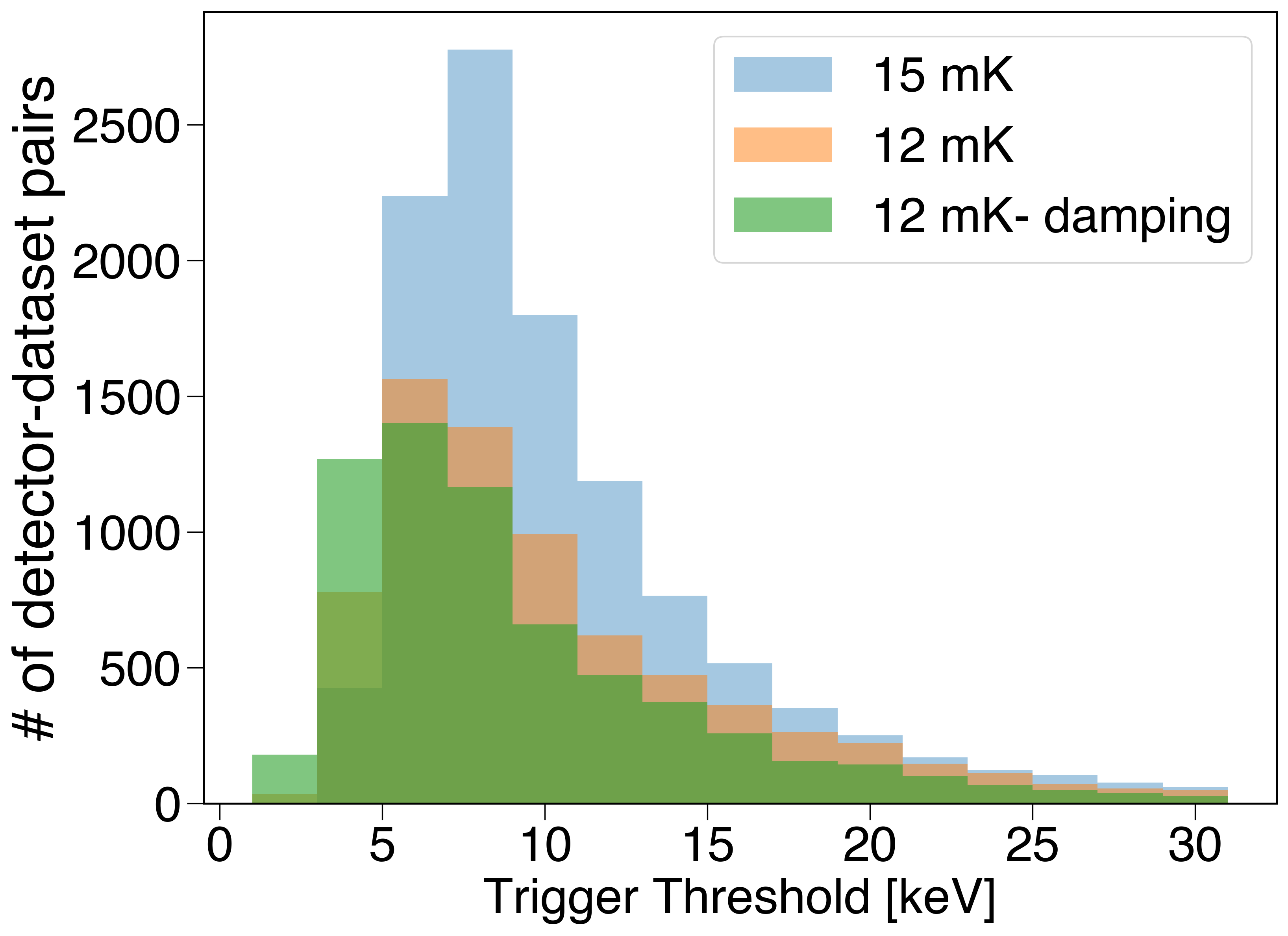}
	\caption{Distribution of the trigger thresholds namely, the energy at which trigger efficiency is 90\%, for the detector-dataset pairs used in the 2 tonne yr low energy dataset. The different colors highlight the pairs belonging to different data taking conditions, i.e. operation temperature at 12 or 15 mK, and employing an optimal vibration damping system configuration. Any distribution refers to all the operated CUORE detectors, but to a different number of datasets, depending on the experimental conditions. }
	\label{fig:OTThresh2TY}
\end{figure}

The distribution of the trigger thresholds by detector-dataset pair is shown in Fig. \ref{fig:OTThresh2TY}. In particular, 59\% of the detector-dataset pairs have a threshold below 10 keV, and 0.9\% have a threshold lower than 3 keV. 
We observe an overall threshold improvement when the detectors are operated at 12 mK, due to an enhanced signal-to-noise ratio. This is further optimized when the oscillation damping system is set to an optimal configuration, that we find to minimize the low-frequency vibration noise. In this configuration, the percentage of detectors with trigger threshold below 3 keV rises to 2.6\%. In contrast, no detectors reach a trigger threshold below 3 keV if the temperature of operation is 15 mK.

\subsection{Tagging of multi-site events}
\label{coincidences}
In most of the physics cases of interest, from 0$\nu\beta\beta$ to dark matter, we search for an energy release in a single CUORE crystal. Therefore, an anti-coincidence cut among CUORE detectors is applied. To identify groups of correlated events, called multiplets, we start by selecting events that occur within a specified time window and spatial radius of a main event. The algorithm then expands the multiplet by searching for additional events, using the same time-space criteria, starting from each event initially included. Finally, it repeats the procedure on all of the events, playing in turn the role of main. Eventually it ensures that multiplets have been built coherently.
The algorithm runs on events satisfying basic quality cuts such as requiring a single pulse in the window and having an energy above a given threshold.

The time window, spatial radius, and energy threshold of the coincidence algorithm can be optimized for different physics searches. First, the energy threshold of the algorithm was set to the trigger threshold for each detector. As a comparison, this is uniformly set to 40 keV for the 0$\nu\beta\beta$ search. To determine the optimal time window for tagging keV-scale multi-site events, we study the signal-to-noise ratio of the M2 Te X-ray peaks while varying the time window across a range of values. We adopt the time window that maximizes the signal-to-noise ratio, namely $\pm$ 15 ms. This window is larger than the one used for higher energy studies in CUORE, namely $\pm$ 5ms \cite{CUOREPRL}, in order to account for the larger spread in the estimate of pulse's peak position in time, due to the lower signal-to-noise ratio at lower energies.

The spatial radius dictates how far apart two crystals can be when building multiplets. In particular, a cut of 150 mm is chosen to ensure only nearby crystals, belonging to the same tower or different ones, are considered to build multiplets of coincident events. In other words, the algorithm searches for coincident events in the eight crystals nearest to the one where the main event occurred. This cut minimizes random coincidences, which are expected to increase at the keV-scale because of a larger rate. 

\subsection{Spurious events discrimination}
\label{spurious}
Approaching the threshold of a detector, we find unwanted features in the energy spectrum. These features are caused by spurious events which mimic physics pulses and can be produced by several effects such as vibrations of detector components, and environmental noise sources such as seismic waves. Although it is not trivial to identify and characterize such spurious sources, we employ dedicated analysis methods to minimize their contribution.

For this purpose, we define two energy intervals, the Region Of Interest (ROI) and the Region Of Reference (ROR): the first determines the lowest energy threshold we aim to analyze, depending on the purpose of the study, while the second is a higher energy interval to be used as a performance comparison. While ROI and ROR are free variables of the algorithm, their specific values for this analysis will be defined in Sec. \ref{dr}.

In the following, as a first selection, we use only detectors having trigger threshold lower than the ROI. An efficient way to reject spurious pulses is to look at their shape, which presents different features than signal waveforms. We explore the performance of several pulse shape variables to find the best one for this discrimination. At first, we study their energy dependence, to make sure they are constant below $\sim$100 keV. Then, we look for the one that optimizes the signal-to-noise ratio of the Te X-rays in the M2 spectrum. We find that the optimal variable is the reduced $\chi^2$ computed between every filtered event and a cubic spline of the filtered average signal pulse used by the OF ($\chi^2_{OF}$). This pulse shape variable has been used as well for low energy studies in CUORE-0 and its predecessors \cite{LowEnergyTechniques,BrancaContour_2020}.

Thus, we aim to remove spurious events by a cut on $\chi^2_{OF}$. Since each detector is affected by noise sources in different ways (because of position in the detector, NTD type, wiring etc.), we expect the optimal pulse shape cut to vary significantly and the cut value to, therefore, be detector-dependent. Moreover, to account for differences in data taking conditions, we treat each dataset separately.

\begin{figure}[!htbp]
    \centering
    \includegraphics[scale=0.2]{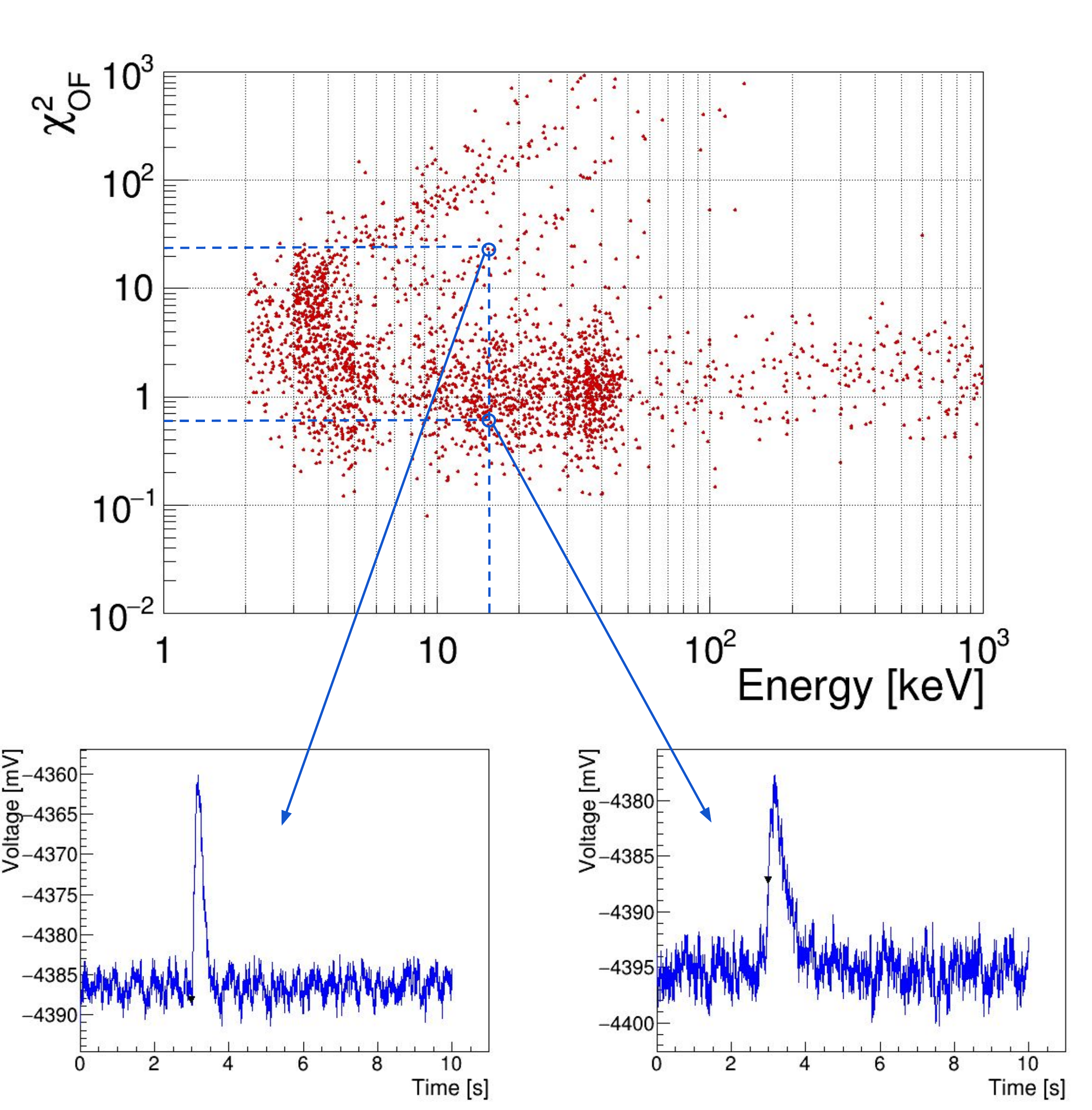}
    \caption{Example of $\chi^2_{OF}$ (Pulse Shape Parameter) as a function of the energy for a single detector, with a $\sim$2 keV trigger threshold, and dataset. We show two pulses having similar energy ($\sim$15 keV), but different $\chi^2_{OF}$. The left pulse belongs to the spurious events band, while the right one to the signal band.}
    \label{examplechitoe}
\end{figure} 

In Fig. \ref{examplechitoe} we show the $\chi^2_{OF}$ distribution over the energy of a single detector and dataset. In this case, we can identify two populations. Physical events, i.e. signal pulses, lay on a constant band at lower $\chi^2_{OF}$ values. Spurious events produce a band at higher $\chi^2_{OF}$ values which increases with energy.

Following this consideration, we apply a pulse shape cut to select events with $\chi^2_{OF}$ lower than a given value. We scan along several cut values to identify the optimal one. For each value we estimate the signal efficiency ($\varepsilon_{PS}$), by fitting the Te X-ray events selected and rejected by the cut simultaneously, and we count the number of events left by the cut in the ROI ($N_B$), assuming they are dominated by physical background or spurious pulses. We identify the optimal cut as the one that maximizes the ratio $\varepsilon_{PS}/N_B$, in order to balance the loss in efficiency with the gain in background level consequent to a stronger cut.

To account for different performance along the CUORE array, we optimize the pulse shape cut independently for each detector. For this purpose, for each detector we define the cut values by looking at the $\chi^2_{OF}$ distribution of a given physical peak (namely, the 511 keV, which is a compromise between low energy and good signal-to-noise ratio). In particular, we divide it in intervals of equal probability (quantiles): the cut values on which we perform the scan are the $\chi^2_{OF}$ values corresponding to the edges of the quantiles, which range from 20 to 90 \% with a 5\% steps. This allows to scan over $\chi^2_{OF}$ values with different scales on different detectors.

\begin{figure}[!htbp]
        \centering
        \includegraphics[scale=0.32]{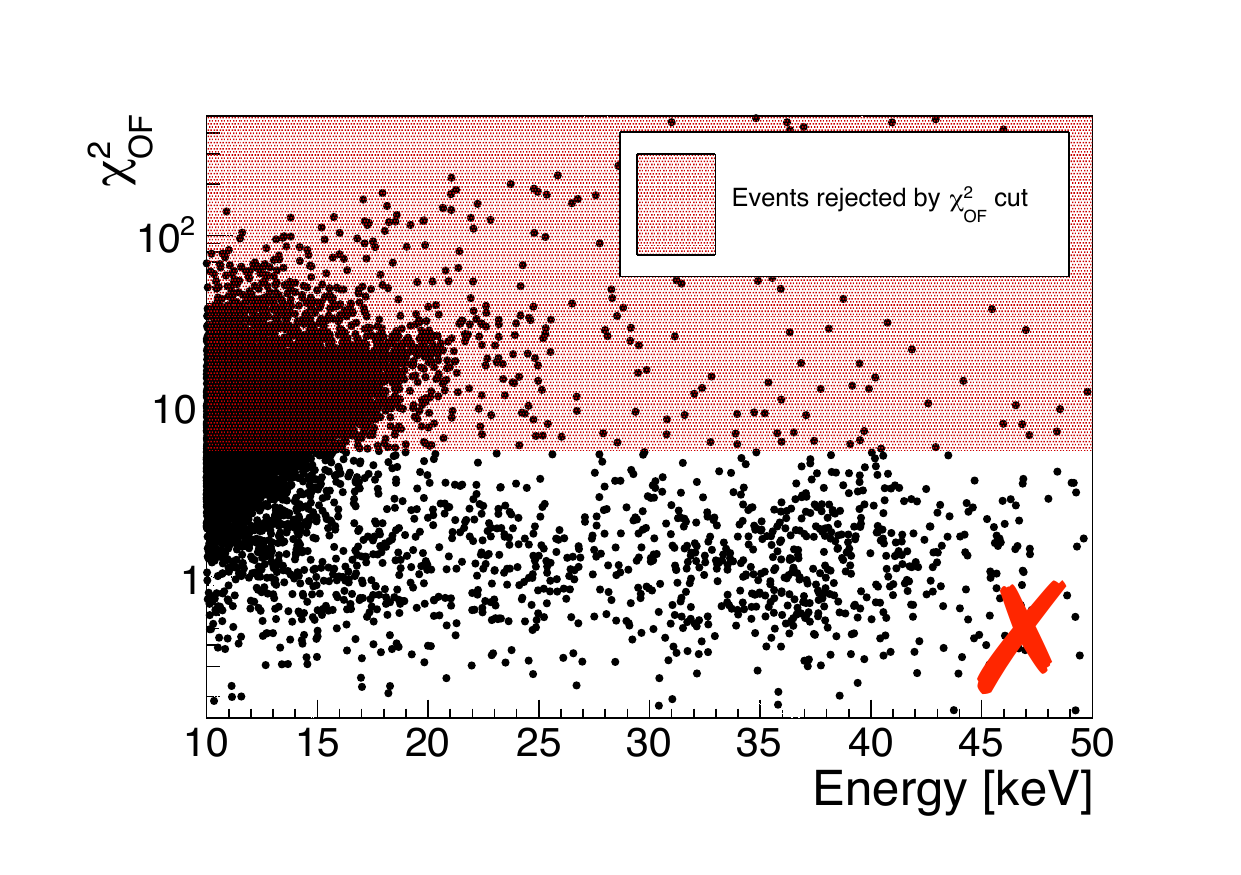}
        \includegraphics[scale=0.32]{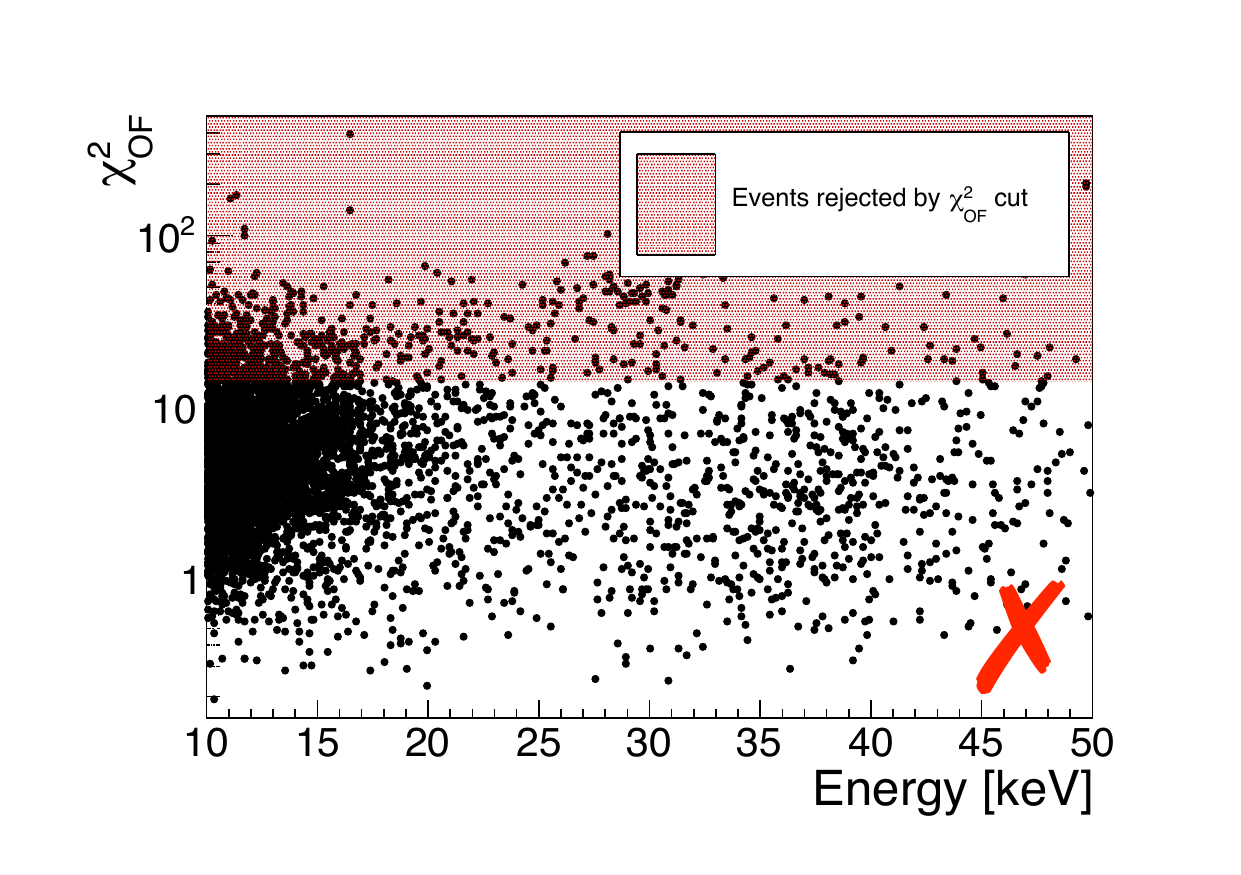}
        \includegraphics[scale=0.32]{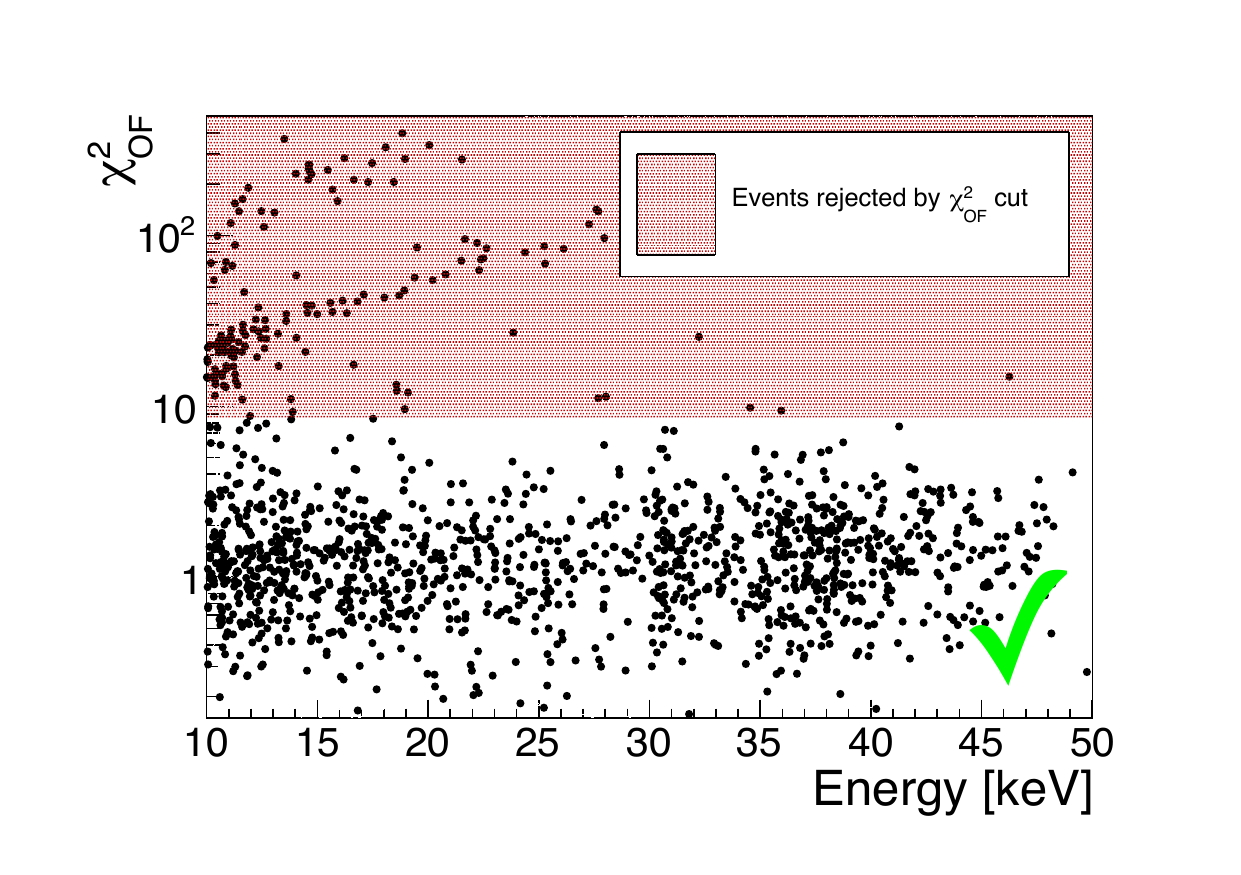}
        \includegraphics[scale=0.32]{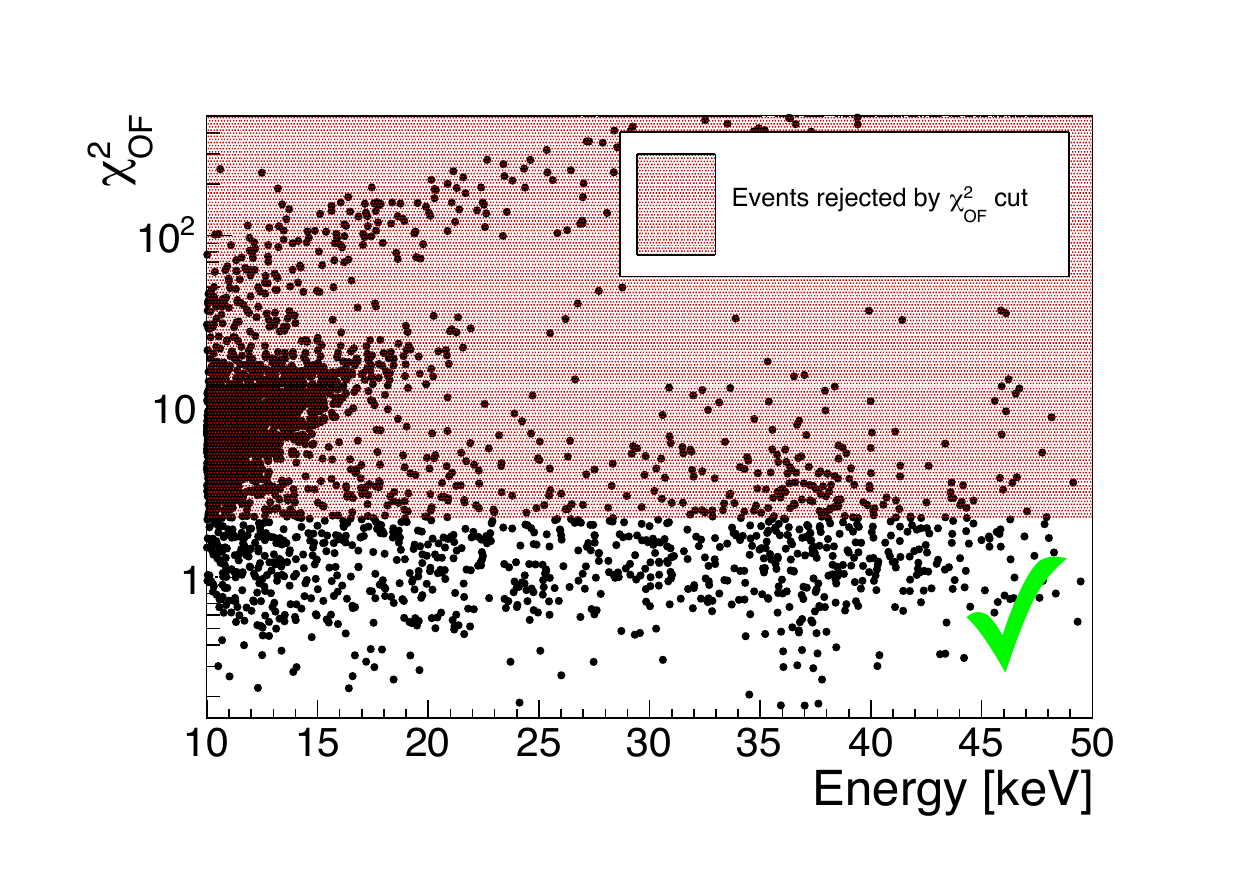}
        \caption{Distribution of $\chi^2_{OF}$ as a function of the energy for four detectors. The red shaded area represents the events rejected by the $\chi^2_{OF}$ cut applied. The first plot from the top shows a spurious band merging with the signal band at low energies. In the second, we do not observe two different bands, but the contamination of spurious events can be identified from the large clustering at lower energy. The two bottom plots report two example in which we expect the pulse shape cut to efficiently reject the spurious band.}
        \label{ch_examples}
\end{figure}

The pulse shape reconstruction by the $\chi^2_{OF}$ variable is inefficient when the energy deposit is close to the noise threshold of the detector. Thus, it is not possible to distinguish signals from spurious pulses below a given energy, which depends on the detector. We report a few examples in Fig. \ref{ch_examples}.
Once the $\chi^2_{OF}$ cut is applied, we can identify detectors which still present unwanted events by looking at two features:

\begin{itemize}
    \item Event Clustering: when the spurious and signal bands merge in the $\chi^2_{OF}$ vs Energy plot, the number of events rises at lower energies. We quantify this effect by building a variable called Purity ($\mathcal{P}$), namely the fraction of signal events in the ROI.
    Assuming a similar density of physical events at different energies, we estimate $\mathcal{P}$ as $N_S/N_B$, where $N_S$ is the number of events in the ROR scaled to the energy width of the ROI. If the ROR is well chosen, after the $\chi^2_{OF}$ cut it contains a pure sample of physics events. The assumption of having the same physical events density at different energies does not affect the detector selection, since we aim to classify and order them from the most to the least contaminated by spurious events.   
    
    \item $\chi^2_{OF}$ median value: despite how the $\chi^2_{OF}$ value of signals and spurious events becomes similar at very low energy, the higher rate of the latter produces a non-uniformity in the relative distribution. Therefore, its median is shifted to higher $\chi^2_{OF}$ values, and this effect is larger as the number of spurious events increases. We define $\Delta \chi^2_{OF}$ as the difference of the medians of the $\chi^2_{OF}$ distribution in the ROI and in the ROR. If the $\chi^2_{OF}$ cut is able to reject the spurious events population, we expect $\Delta\chi^2_{OF}$ to approach 0.
    \end{itemize}
    
In the following, we adopt a ROR between 30 and 50 keV. This choice is led by the necessity to maintain consistent pulse shape characteristics and event rates with respect to the ROI. 
The second requirement is no longer satisfied above $\sim$50 keV due to an observed difference by about an order of magnitude of the background level, as better outlined in Sec. \ref{spectra}.

We aim to determine if the $\chi^2_{OF}$ cut is sufficient to reject spurious events in the ROI while we expect it to completely remove these events from the ROR.

For this purpose, we compute $\mathcal{P}$ of all the detectors, obtaining a distribution from 0 to 1. Then, we conduct a scan on this variable by using steps of 0.05. First, we select the detectors with $\mathcal{P}$ higher than 0, and we compute the resulting exposure (mass $\times$ livetime, $M \Delta T$) and $N_B$. Then, at the following step, we select detectors with $\mathcal{P}$ higher than 0.05 and we repeat the computation of $M \Delta T$ and $N_B$. We iterate the procedure until the last step and we define the cut value on $\mathcal{P}$ based on the one that maximizes $M \Delta T/N_B$, to balance the loss in exposure with the gain in background achieved from selecting a smaller subset of better-performing detectors. 

Similarly, we perform a scan on $\Delta \chi^2_{OF}$ starting from the maximum value among all the detectors and ending at 0. In this case, we compute $M \Delta T/N_B$ with the subset of detectors having lower $\Delta \chi^2_{OF}$. Again, we select the detectors based on the value which maximizes $M \Delta T/N_B$.

\begin{figure}[!htbp] 
    \centering
    \includegraphics[scale=0.4]{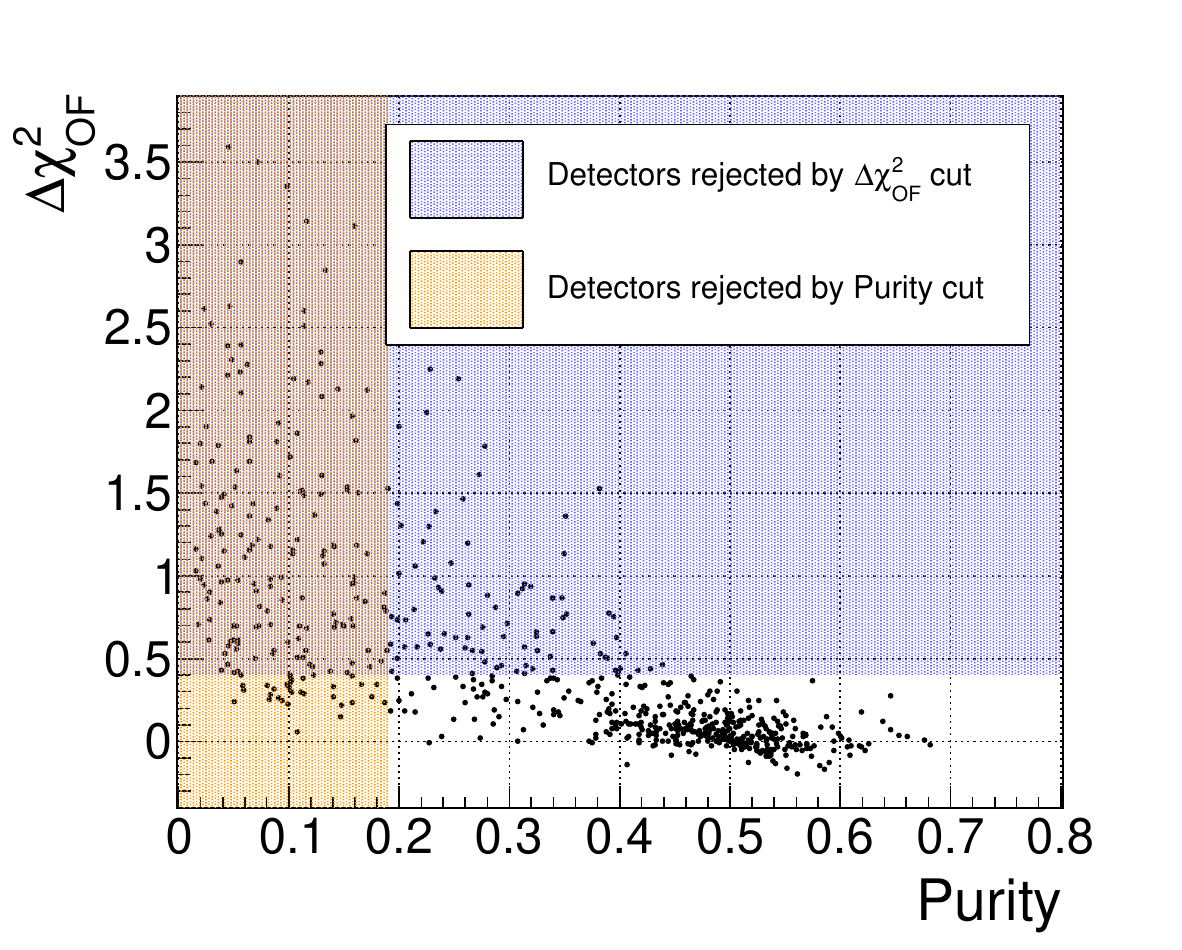}
    \caption{The plot shows the anti-correlation of the two variables used to discriminate spurious pulses. Each data point in the plot is a CUORE detector from an example dataset. Good performing detectors have low $\Delta \chi^2_{OF}$ and high purity. The orange and blue shaded regions represent the regions rejected by the optimized cut on Purity and $\Delta \chi^2_{OF}$ to select the best performing detectors at a given energy threshold (10 keV in this example).}
    \label{pandchi}
\end{figure}

The intersection of the detectors selected by the cut on $\mathcal{P}$ and $\Delta \chi^2_{OF}$ represents the final subset. As shown in Fig. \ref{pandchi}, the 2 quantities are anti-correlated, meaning that by selecting high $\mathcal{P}$ detectors we are implicitly selecting low $\Delta\chi^2_{OF}$ and viceversa. This method ultimately selects a small subset of detectors where the 
$\chi^2_{OF}$ cut ensures a single, constant, and uniformly-populated band of low-energy events in the 
$\chi^2_{OF}$ vs. energy plane, as shown in Fig. \ref{ch_examples}, confirming the expected behavior of physical events.
        
\section{Data Selection Results}
\label{dr}
We apply the previously-described methods on the 2 tonne yr CUORE data release using two different ROIs: [10,20] keV, of interest for solar axions searches \cite{AxionM1CUORESearch,axionreview,Li_2015,Li_2016}, and [3,10] keV. The latter is the lowest accessible energy region of CUORE since no detectors achieved a trigger threshold lower than 2 keV. Moreover, it is of interest to study the excess of events at 4.7 keV already observed in past CUORE demonstrators \cite{lowEspectrum}. The $\chi^2_{OF}$ cut optimization and the detectors selection is repeated for each chosen ROIs.
We achieve a total of 691 and 11 kg yr TeO$_2$ exposure for detectors available down to 10 keV and 3 keV, respectively. This amounts to 34\% and 0.5\% of the total collected exposure, respectively. Considering only 12 mK datasets, the exposure at 3 keV reaches 1.2\% of the total one. 

\begin{figure}[!htbp]
    \centering
    \includegraphics[scale=.2] {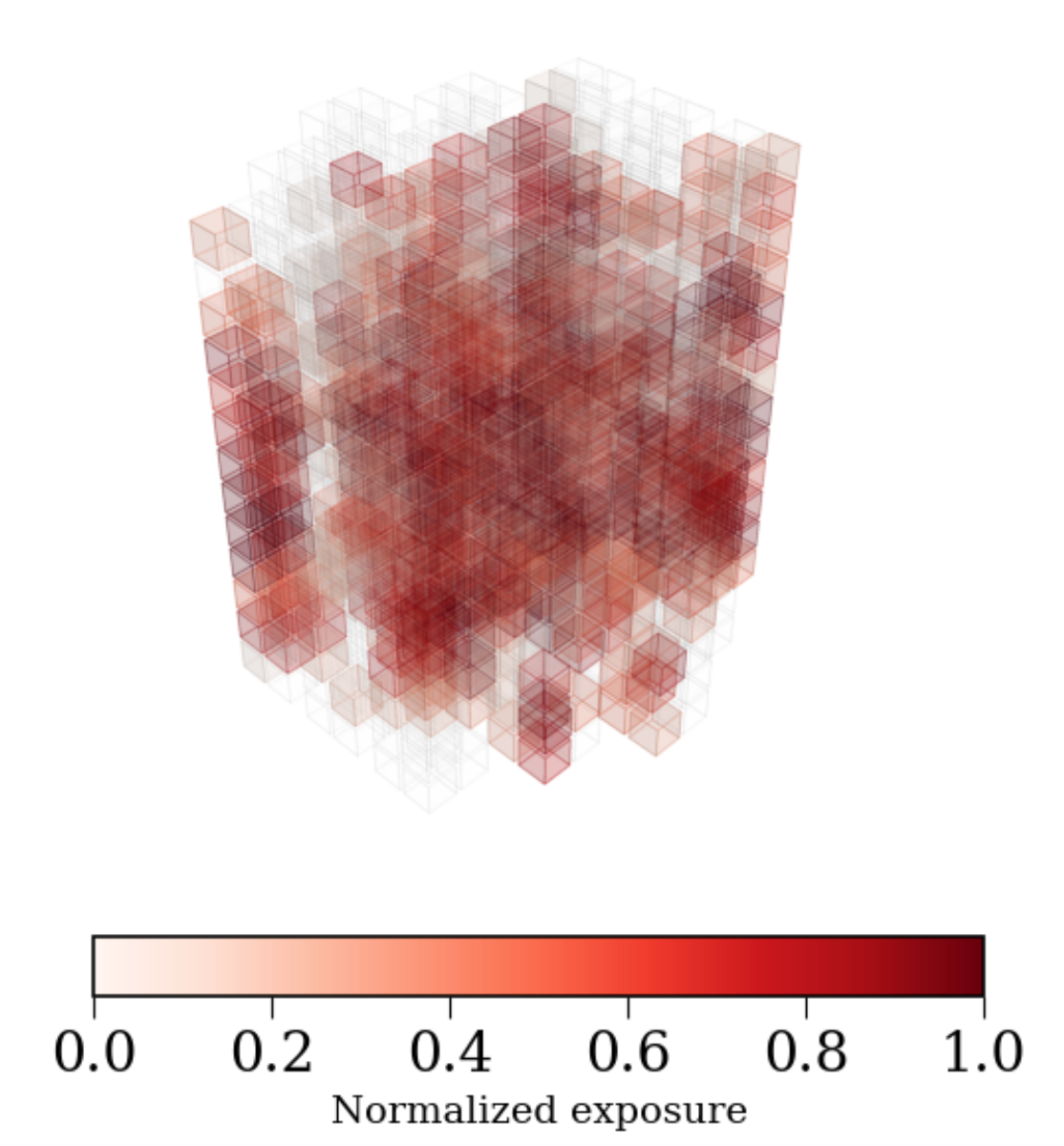}
    \includegraphics[scale=.2] {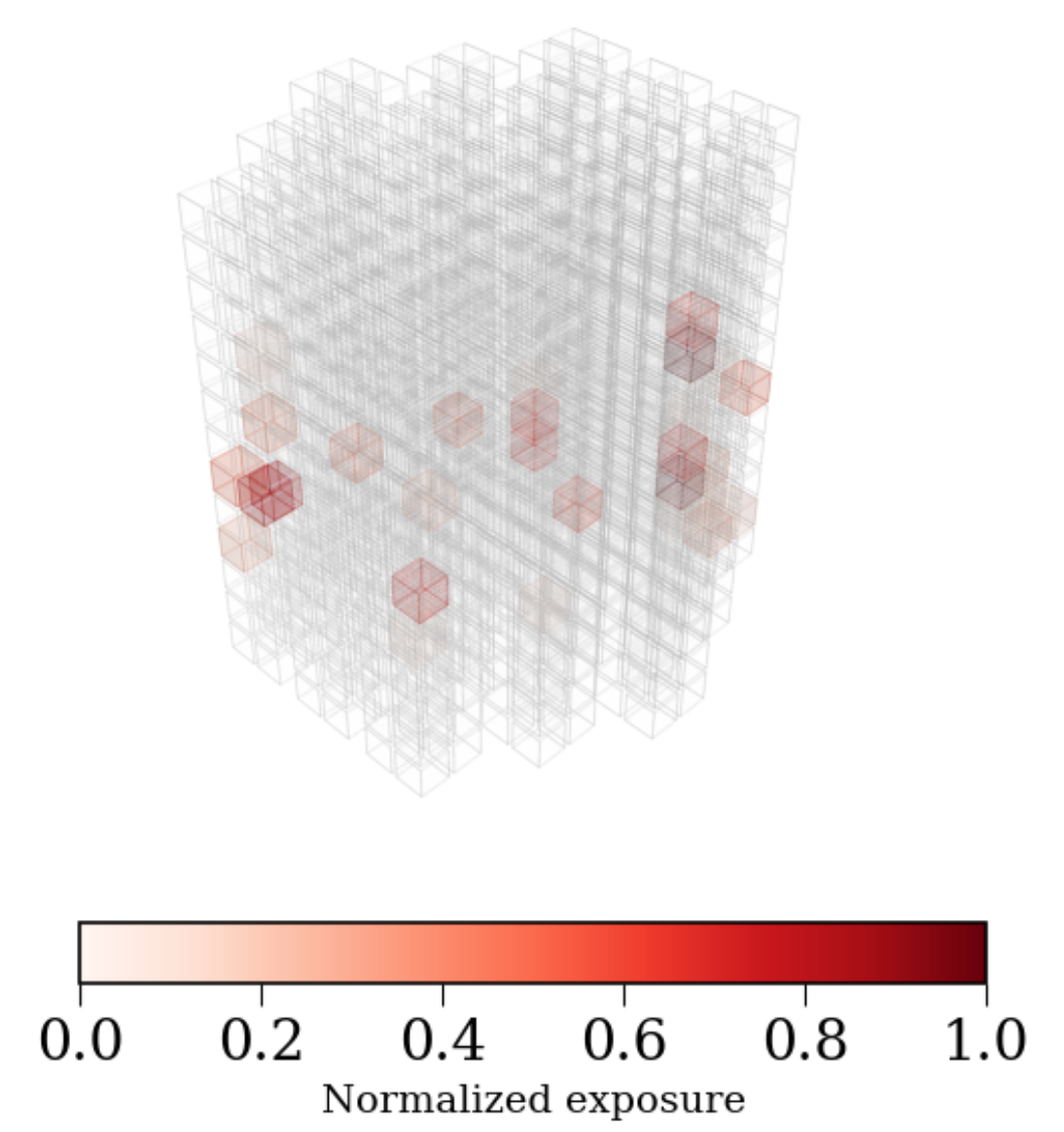}
    \caption{Detectors selected at 10 keV (top) and 3 keV (bottom) in the CUORE array. The color scale highlights the exposure available for each of them, normalized to the detector with the highest. This depends on the number of datasets in which a given detector was selected and the number of high-rate runs we rejected for each of them.}
    \label{fig:selecteddetectors}
\end{figure}

The detectors selected for the two energy thresholds in the CUORE array are shown in Fig. \ref{fig:selecteddetectors}, highlighting the fraction of exposure available for each of them.

The developed detector selection method excludes the majority of the upper floor detectors (see Fig. \ref{fig:selecteddetectors}). Namely, among the detectors selected down to 10 keV, 24\%, 44\% and 32\% belong to the top 5, middle 4 and bottom 4 floors, respectively. Considering detectors selected down to 3 keV, the difference is even more evident because of the stricter selection: only 0.7\% belong to the 5 top floors, while 57.7\% belong to the middle 4 floors and 41.6\% to the 4 bottom ones. The preference for detectors in the middle and bottom floors of CUORE suggests that these floors are less affected by different types of vibrational noise. 
This is supported by evidence that lower floors are less sensitive to sub-Hz noise induced by microseism events originating from seastorm activity reported in Ref. \cite{seastormscuore}.

Moreover, the low energy selection highlights the better performance of detectors equipped with a specific type of NTD. In particular, the NTDs of CUORE were made in three batches that differed slightly in irradiation, leading to differences in properties. Each tower in the setup exclusively utilizes NTDs from a single batch, ensuring consistency within that tower. Most of the selected detectors, namely 47\% and 85\% at 10 and 3 keV, respectively, belong to towers with the least resistive of the NTD batches \cite{IreneNThesis}. The lower resistivity leads to a lower Johnson noise, which contributes to a better baseline resolution and energy threshold. 

\begin{figure}[!htbp]   
\centerline{
    \includegraphics[scale=.2]{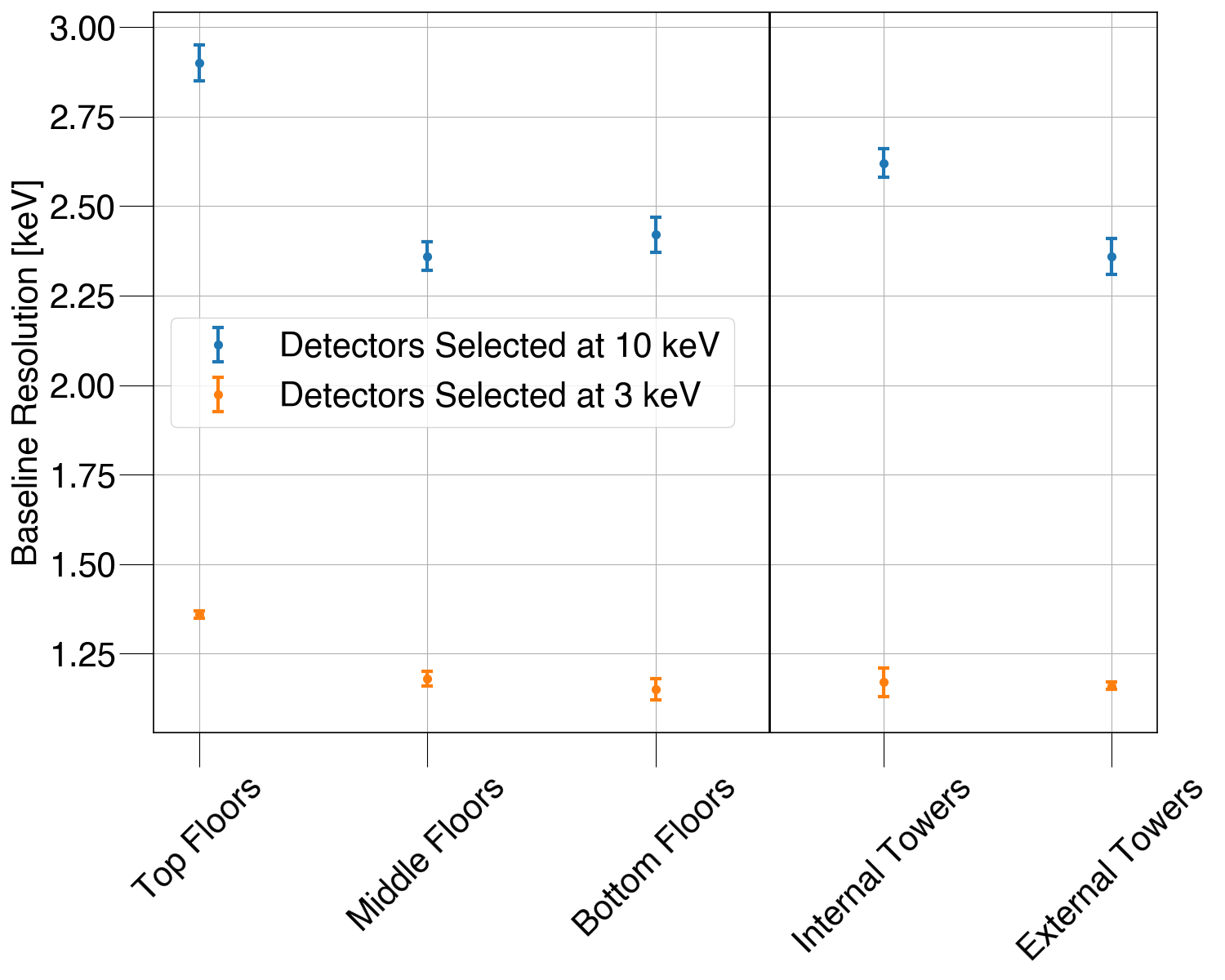}}
    \caption[Baseline resolution by geometry]{The average baseline resolution of the detectors selected at 10 keV and 3 keV is reported for different portions of the CUORE array, namely three groups of floors, internal and external towers. The uncertainty is taken as the standard deviation of the mean among the selected detectors of a given category. This highlights the differences in energy resolution observed especially in the top floors.}
    \label{fig:baselinegeometry}
\end{figure}

Finally, there is not a significant preference for internal with respect to external towers among the selected detectors both at 10 keV and 3 keV. In particular, 57\% and 26\% of the selected detectors at 10 and 3 keV, respectively, belong to internal towers. This means that a potential lower physical background, as expected for inner towers, does not affect the detector selection algorithm. However, we observe a difference ($\sim$10\%) between internal and external towers in terms of baseline resolution among the detectors selected at 10 keV, as shown in Fig. \ref{fig:baselinegeometry}. The difference is mitigated when applying the stricter selection at 3 keV. This reflects the fact that most of the best performing NTDs are in the external towers. Similarly, we find a different baseline resolution comparing different floors (see Fig. \ref{fig:baselinegeometry}), likely due to the different sensitivity to vibrations.

\begin{figure}[!htbp]

    \includegraphics[scale=.27]{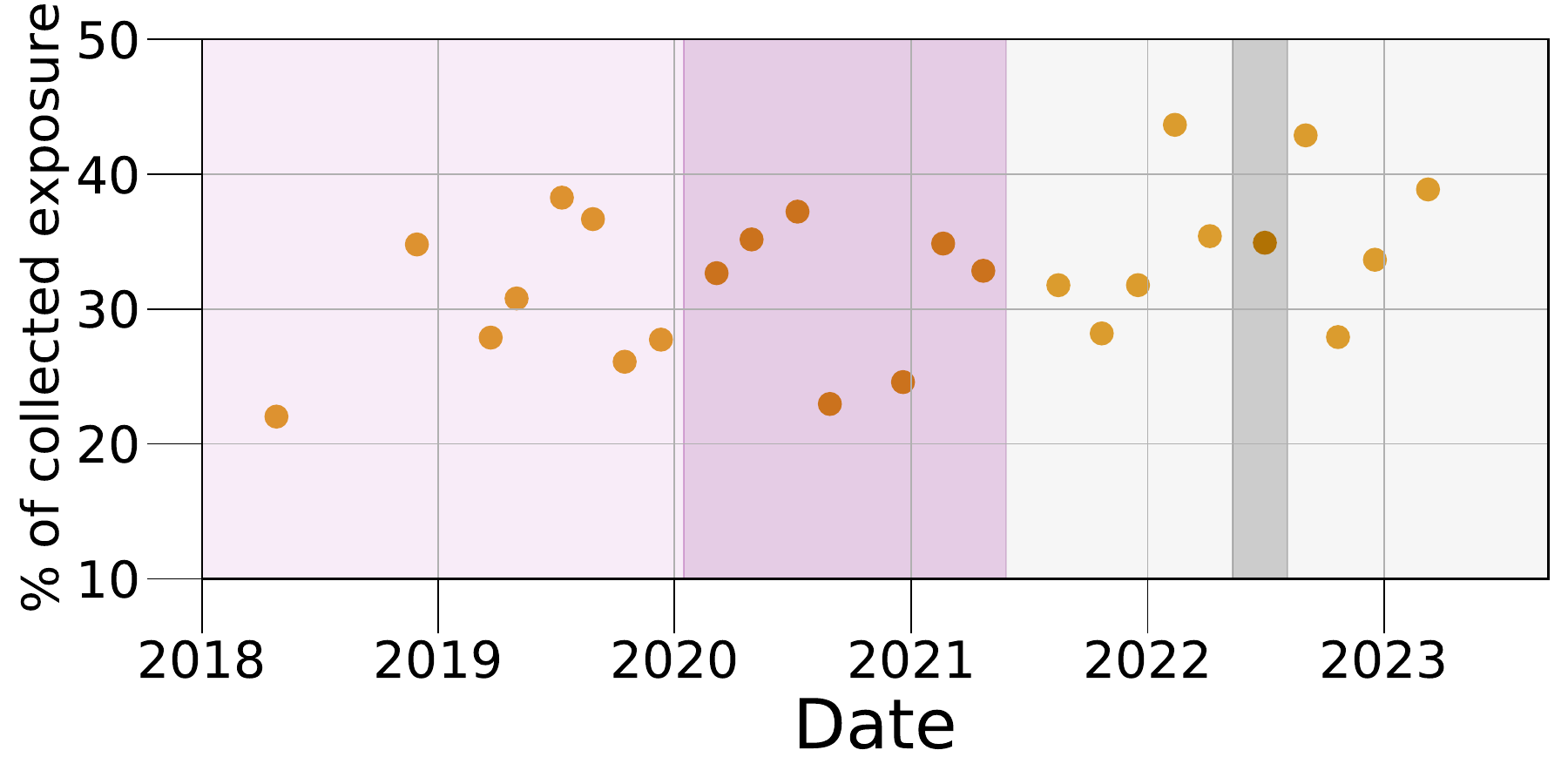}   \includegraphics[scale=.27]{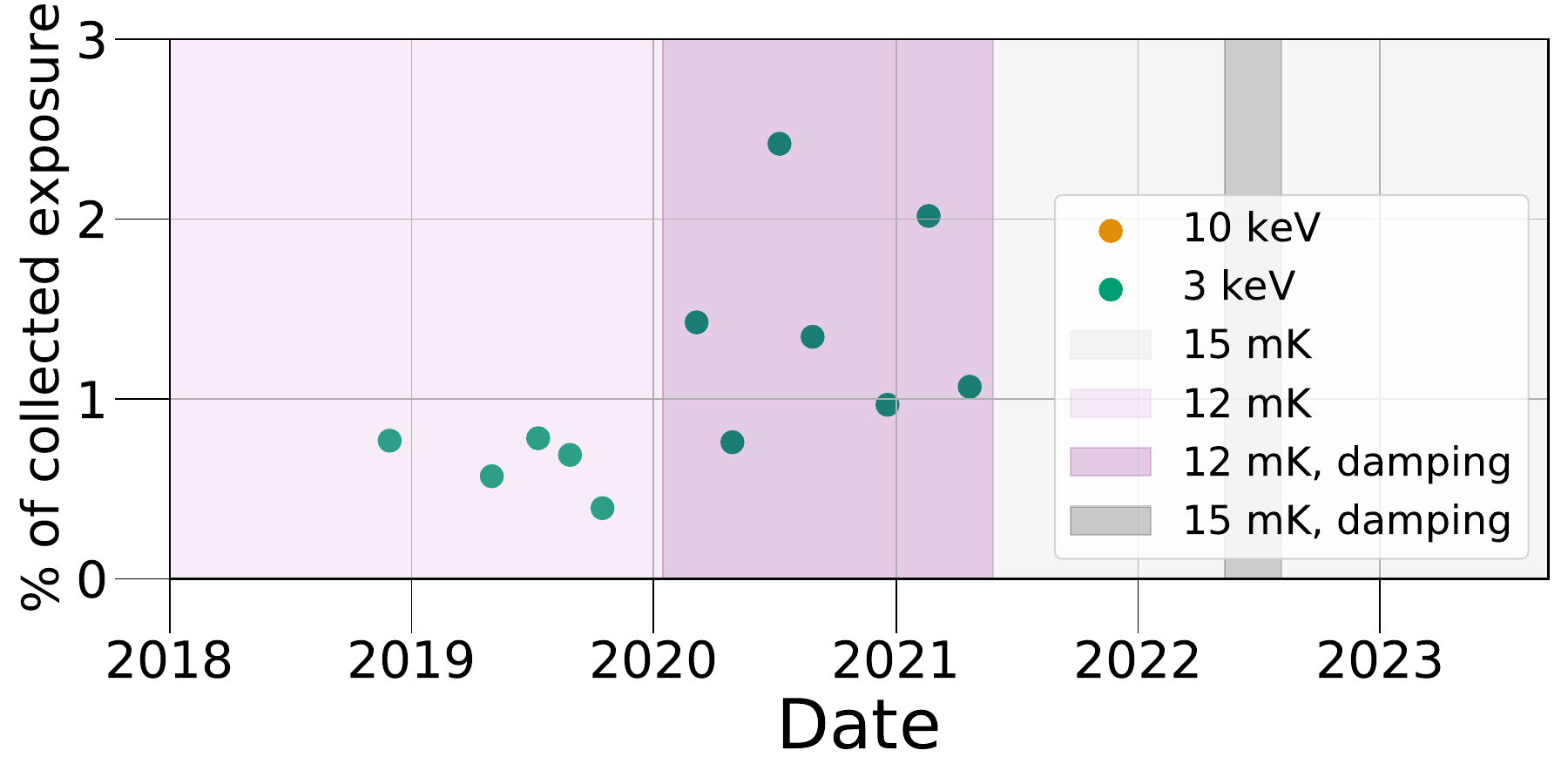}
    \caption{Percentage of exposure selected at 10 keV, on the top panel, and at 3 keV, on the bottom panel, by datasets. The color bands indicate different data taking conditions.}
    \label{fig:LowEExposure}
\end{figure}

\begin{figure}[!htbp]  
    \includegraphics[scale=.06]{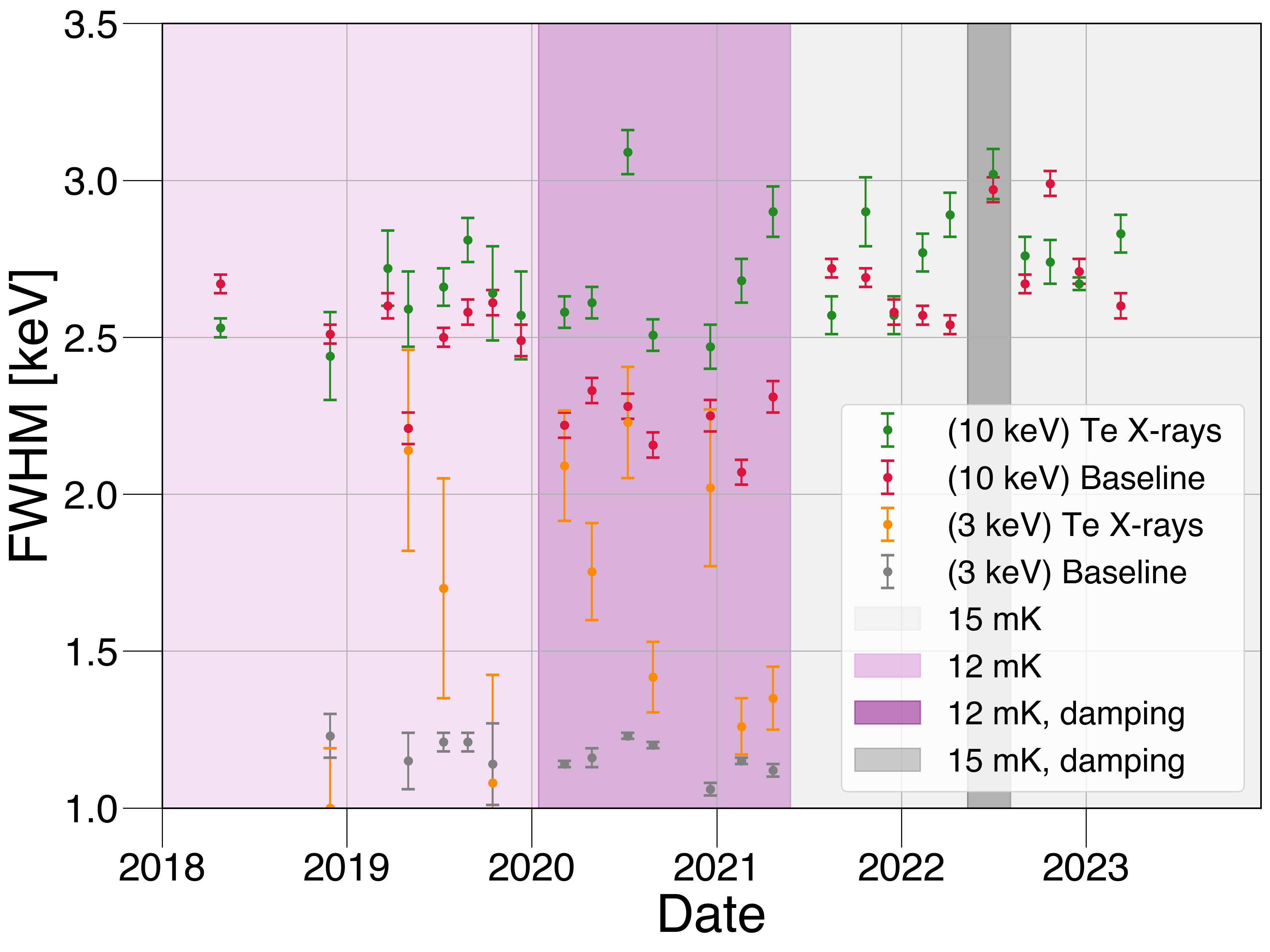}
    \caption[Energy resolution of 2 tonne yr low-energy reprocessing]{Energy resolution at Te X-rays and at the baseline of each dataset. The uncertainty reported is estimated from the fit in the case of Te X-rays resolution, and it is the standard deviation of the mean among the detectors of a dataset in the case of baseline resolution. Overall poorer energy resolution was observed in datasets with non-optimal damping configuration and higher operation temperature.}
    \label{fig:LowEEnergyRes} 
\end{figure}

The selected exposure, or number of detector-run pairs, by dataset has various trends when compared to data taking conditions, as mentioned in Sec. \ref{intro}, such as detector temperature and vibration damping configuration, as shown in Fig. \ref{fig:LowEExposure}. The operation temperature heavily affects the exposures available at 3 keV, as no detectors pass selections at this energy in datasets operated at a warmer 15 mK temperature. The time periods labeled as ``damping" in Fig. \ref{fig:LowEExposure} and \ref{fig:LowEEnergyRes} correspond to an optimal configuration of the custom vibration isolators made by Minus K Technology \cite{minusk}. This system is located under the Y-beam that supports the CUORE Top Support Plate of the detector and it is meant to mitigate sub-Hz vibration noise, with a cut-off at 0.5 Hz\cite{CUOREInfrastructure}. In particular, we find that a specific configuration operated over a few datasets is beneficial to mitigate the noise at very low energy, leading in a broader subset of detectors usable down to 3 keV.

The same trend along the three phases of data taking described in Sec. \ref{intro} holds when considering energy resolution, as it improves with colder temperatures and the optimal damping configuration. We compute the baseline resolution by fitting to a Gaussian function the energy distribution of randomly triggered noise events with no pulses in the window, and getting the FWHM resolution. In the optimal data taking configuration we can appreciate the larger Te X-rays peak width due to its composition made of multiple lines. 

In Fig. \ref{fig:LowEEnergyRes} we report the average baseline resolution among the detectors selected down to 10 and 3 keV for each dataset. As a comparison we also report the energy resolution estimated by fitting Te X-rays in calibration data.

\section{Efficiencies}
\label{eff}
On each dataset, we evaluate efficiencies for anti-coincidence and pulse shape (PS) cuts, and for reconstruction on the detectors selected at 10 and 3 keV. 

The reconstruction efficiency is the product of trigger detection, energy reconstruction, and pile-up rejection efficiencies. Together, these are the probability that a signal is triggered, is reconstructed with the correct energy, and is the only pulse in the signal window. In the standard CUORE data processing, reconstruction efficiencies are calculated from heat signals injected through the Si chips, in the following referred to as ``pulsers". The frequency and amplitude of these pulsers are known and adjustable. These are the same kind of pulsers used for thermal gain correction. 

Throughout CUORE data taking, the keV-scale injected pulsers are programmed in different ways. For approximately the first tonne yr of exposure collection, they were acquired during dedicated runs taken between datasets and lasting up to 48h. For most of the second tonne yr, instead, they were injected during physics data taking.

Unfortunately, in a few datasets, the data taking conditions didn't allow the acquisition of pulser events at all. In this case, an average efficiency from datasets acquired in the same experimental configuration is used. 

\begin{figure}[!htbp]
\centering
\includegraphics[scale=.07] {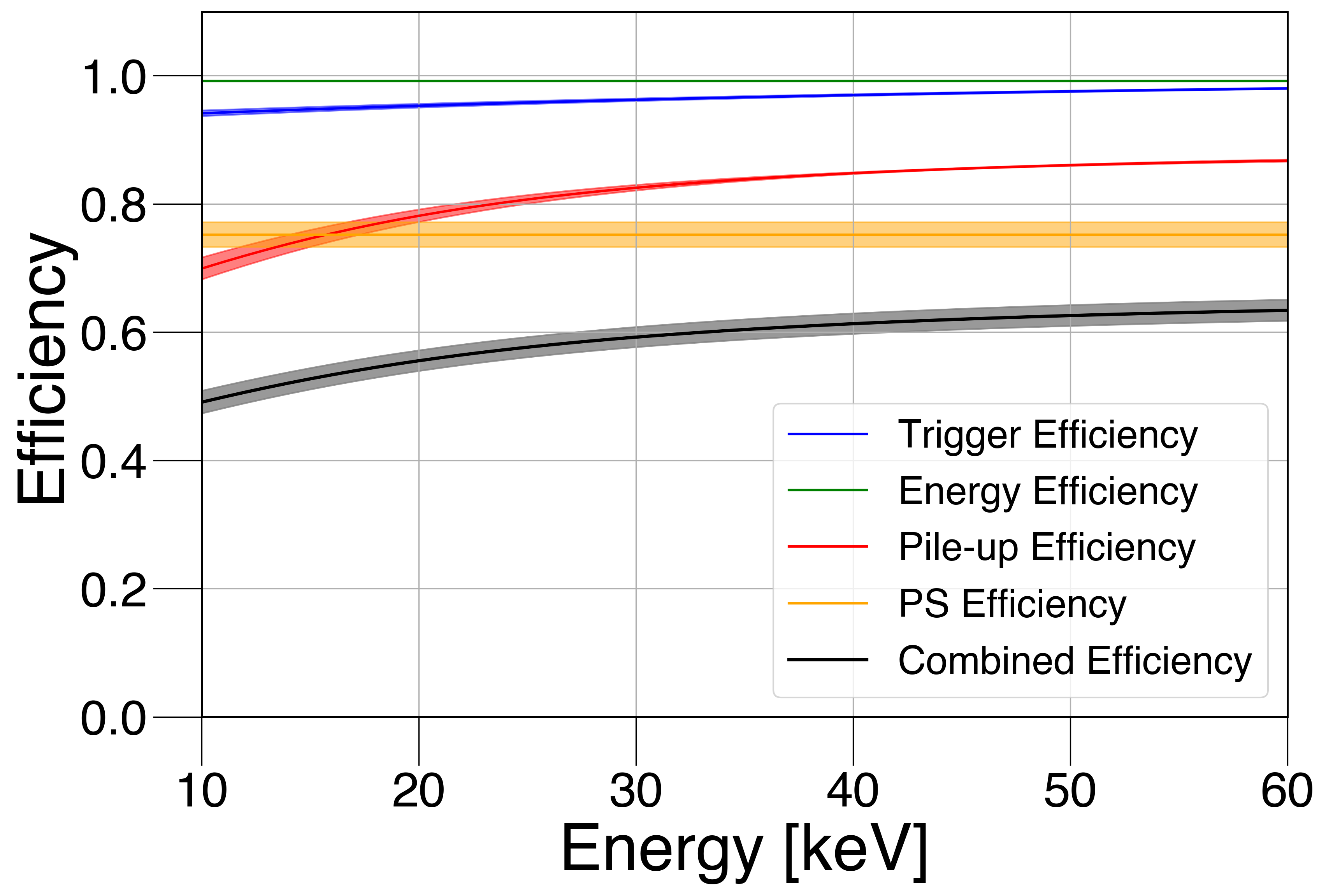}
\includegraphics[scale=.07] {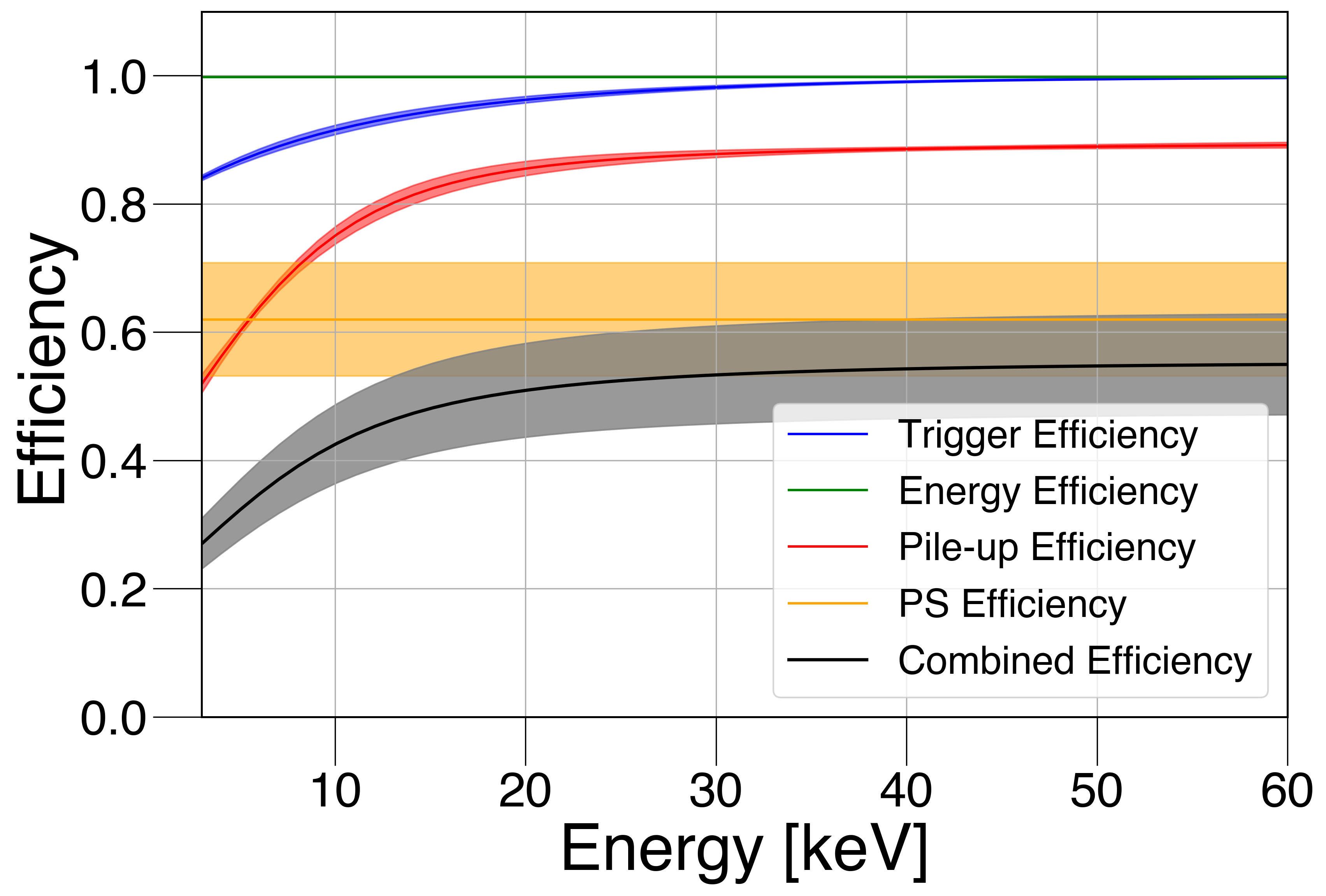}
\caption{\label{EffCombined3_10keV} The curves show the efficiencies (averaged among all the datasets) as a function of the energy and their combination for the detectors selected at 10 keV (top panel) and  3 keV (bottom panel) in the 2 tonne yr low energy datasets. The band around each efficiency curve represents $\pm$ 1 $\sigma$ combined from the uncertainty on each dataset. }
\end{figure}

In previous CUORE analyses, only pulsers with an average energy of $\sim$3 MeV, were used to calculate these efficiencies \cite{CUOREPRL,nature,cuorecollaboration2024nuhuntingseedmatterantimatter}. We adapt the same algorithm to run on lower energy pulser events to evaluate the near-threshold reconstruction efficiencies and evaluate their energy dependence. In particular, we evaluate the reconstruction efficiencies on a dataset basis, for every group of pulser events at decreasing energy  from $\sim$350 keV down to $\sim$5 keV. Then, for each dataset, we fit each of the reconstruction efficiencies as a function of the energy to a logistic function. The resulting curve for each efficiency is averaged among all the datasets weighted by exposure. Finally, the overall reconstruction efficiency is produced by analytically multiplying all the exposure-weighted average efficiencies. The result is shown for the detectors selected at 10 keV and 3 keV in Fig. \ref{EffCombined3_10keV}.

The anti-coincidence and $\chi^2_{OF}$ cut efficiencies are evaluated by using $^{40}$K and Te X-ray peaks, respectively, before and after applying the corresponding cuts. 

We assume the $^{40}$K line is a pure M1 event, and we consider the presence of the previously described $\sim$3.2 keV X-ray to be negligible since its emission intensity is as low as $\sim$0.07\% \cite{nucleardata}. Without knowledge of known physical structures in the spectrum to evaluate these efficiencies, we assume them to be constant along the energy scale of interest for this work. The anti-coincidence efficiency is found to be compatible with 1 in all cases and so it doesn't affect estimations of the overall efficiency.

\section{Low Energy Spectra}
\label{spectra}
\begin{figure}[!htbp]
\centering
\includegraphics[scale=.4] {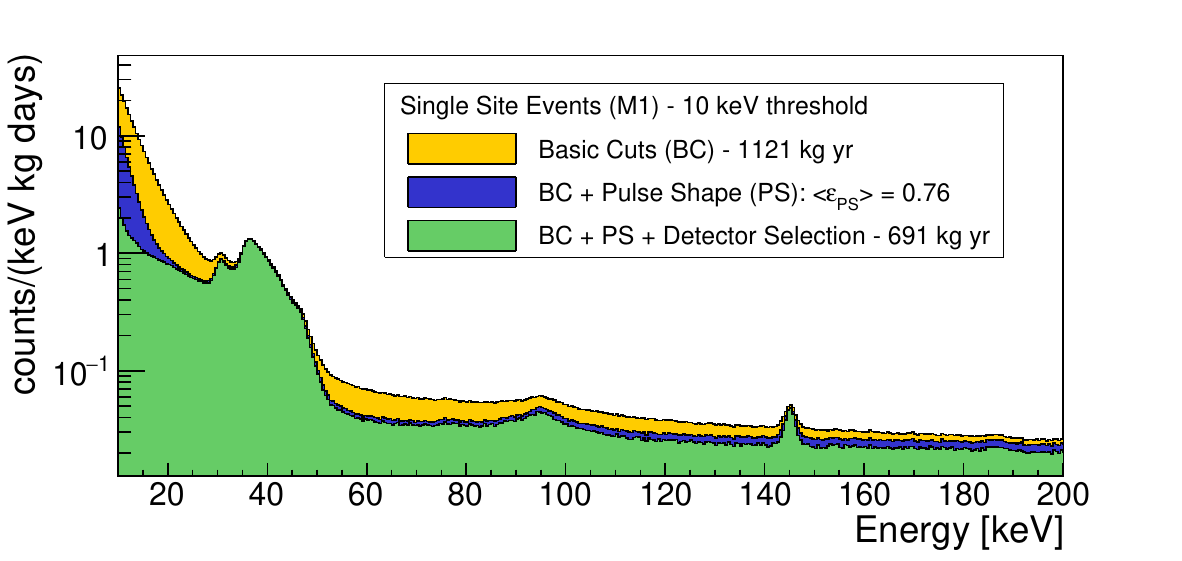}
\qquad
\includegraphics[scale=.4] {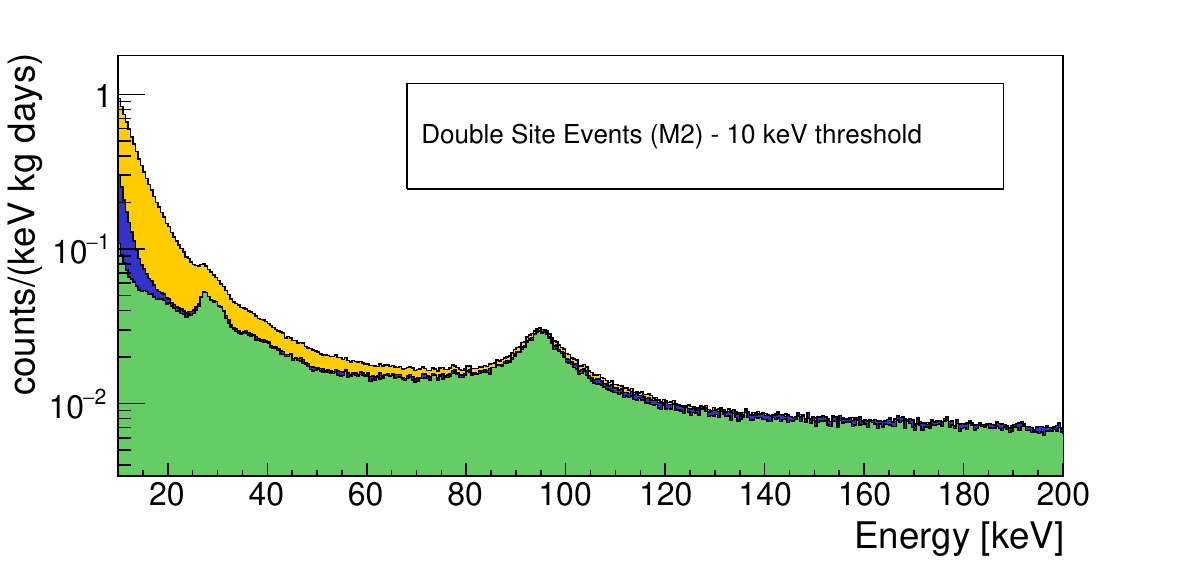}
\caption{Energy spectra with multiplicity equal to 1 and 2, on top and bottom, respectively, up to 200 keV using the data selection optimized for the threshold of 10 keV. The different colors highlight the spectra after applying the requirement that the trigger threshold is less than 10 keV (in yellow), the detector dependent cut on $\chi^2_{OF}$ (in blue), and the detector selections (in green). The last two are weighted by the $\chi^2_{OF}$ cut efficiency, which is on average 76\%. The final spectrum presents an exposure of 691 kg yr.}
\label{fig:LowE10keV}
\end{figure}

\begin{figure}[!htbp]
\centering
\includegraphics[scale=.4] {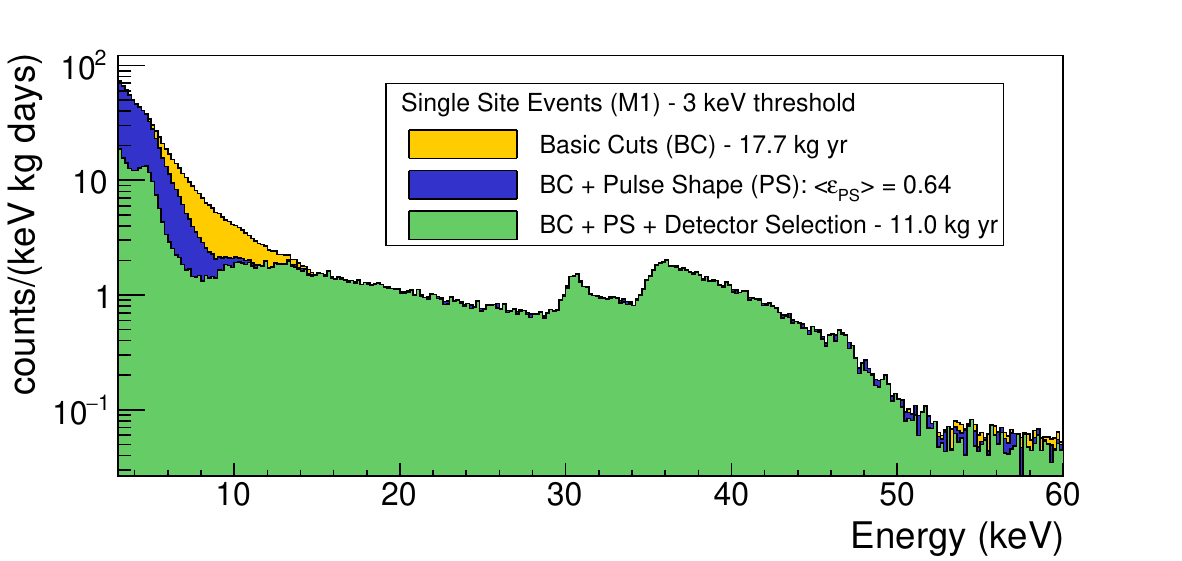}
\qquad
\includegraphics[scale=.4] {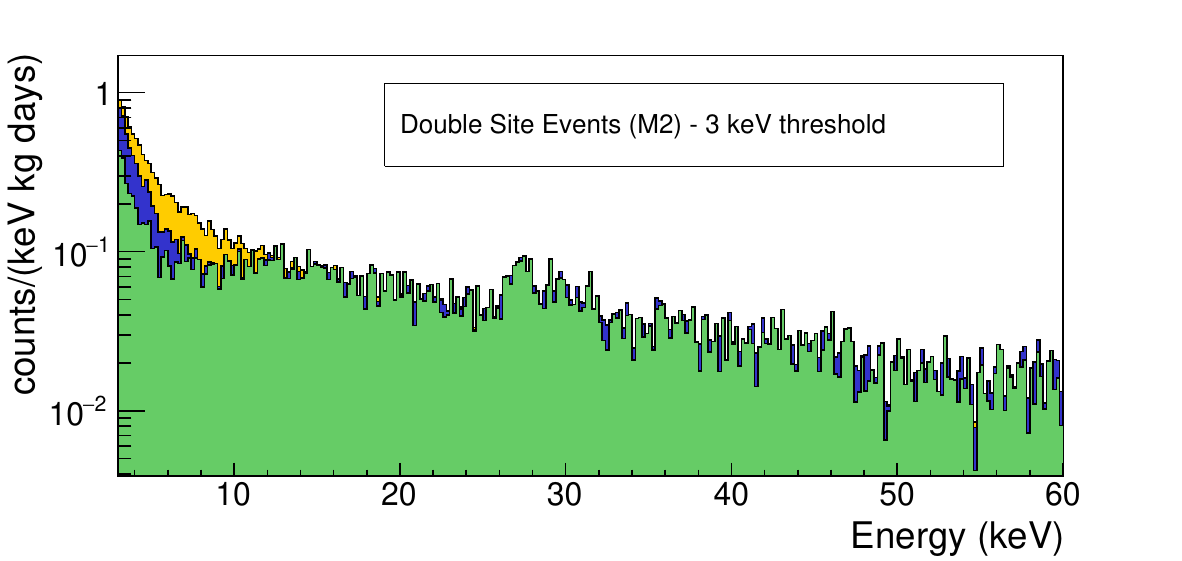}
\caption{Energy spectra with multiplicity equal to 1 and 2, on top and bottom, respectively, up to 60 keV using the data selection optimized for the threshold of 3 keV. The different colors highlight the spectra after applying the requirement that the trigger threshold is less than 3 keV (in yellow), the detector dependent cut on $\chi^2_{OF}$ (in blue), and the detector selections (in green). The last two are weighted by the $\chi^2_{OF}$ cut efficiency, which is on average 64\%. The final spectrum presents an exposure of 11 kg yr.}
\label{fig:LowE3keV}
\end{figure}

In Fig. \ref{fig:LowE10keV} and \ref{fig:LowE3keV} we show the exposure-weighted low energy background spectrum of the 2 tonne yr of CUORE data taking down to 10 and 3 keV, respectively. 

The initial cuts require the trigger threshold to be less than the chosen lower edge, either 10 or 3 keV (spectrum in yellow). Next, we apply the detector-dependent cut on $\chi^2_{OF}$ (in blue), and lastly the detector selections (in green). The blue and green spectra are weighted by the $\chi^2_{OF}$ cut efficiency. In this way, it is possible to compare the background level reduction due to the rejection of spurious pulses before and after the cut.  The $\chi^2_{OF}$ cut and the detector selection show the greatest difference from basic cuts below 30 keV, reducing the background by an order of magnitude. This region is the most interesting for multiple planned and potential physics searches.

The result of this work highlights the main spectral features of CUORE at low energy which are summarized in Table \ref{table:spectrumfeature}.

\begin{table}[!htbp]
	\centering

	\begin{tabular}{ccc}
		\hline
		\hline
		\textbf{Energy [keV]} & \textbf{Multiplicity} & \textbf{Origin} \\
		\hline
		$\sim$ 27, 31 & M2 & Te X-rays \\

		$\sim$46 & M1 & $^{210}$Pb $\beta+\gamma$ \\

		$\sim$90 & M1, M2 & $^{210}$Po nuclear recoil \\

		$\sim$145 & M1 & $^{125m}$Te  \\
		\hline
		\hline
	\end{tabular}
	
	\vspace{0.8em}
	
	\begin{tabular}{ccc}
		\hline
		\hline
		\textbf{Energy [keV]} & \textbf{Multiplicity} & \textbf{Rate [counts/(kg d)]} \\
		\hline
		$\sim$4.7 & M1 & 19.1 $\pm$ 1.0 \\

		$\sim$10 & M1 & 0.66 $\pm$ 0.07 \\

		$\sim$13 & M1 & 0.67 $\pm$ 0.07 \\

		$\sim$31 & M1 & 1.35 $\pm$ 0.04 \\

		$\sim$36 & M1 & 8.3 $\pm$ 0.3 \\
		\hline
		\hline
	\end{tabular}
	
	\caption{Summary of the main features of CUORE energy spectrum up to 200 keV.
		The top table reports structures of known origin, while the bottom one those which are currently under further investigation.}
	\label{table:spectrumfeature}
\end{table}

Starting from the higher energies of the spectrum in Fig. \ref{fig:LowE10keV}, we observe a monochromatic peak at $\sim$145 keV, likely due to the $\beta$-decay of the contaminant $^{125}$Sb \cite{datadriven}, and the $^{125m}$Te de-excitation.

The $\sim$90 keV events likely originate from nuclear recoil from $^{210}$Po $\alpha$-decay on crystal surfaces. This explanation gains support from the fact that the peak structure is more pronounced in the M2 spectrum compared to the M1, suggesting a surface contamination. The hypothesis has been verified by looking at the energy of M2 coincident events, which peaks at $\sim$5.3 MeV, the energy of the $\alpha$ emitted by the $^{210}$Po.

The spectrum shows an increase in the background level moving down from $\sim$50 keV to $\sim$30 keV by about an order of magnitude, featuring two peaks at $\sim$31 and $\sim$36 keV.

The nature of the $\sim$36 keV structure is still unclear, but it has been observed in other measurements using TeO$_2$ cryogenic calorimeters \cite{LowEnergyTechniques,lowEspectrum}. It was originally ascribed to the $^{210}$Pb in Ref. \cite{123Te2003}, where this structure was observed together with a $\gamma$ peak at $\sim$46 keV. Nevertheless, both CUORE-0 and the data discussed in this work exhibit a kink at $\sim$46 keV in the spectrum, rather than a monochromatic peak, indicating the presence of additional continuous contributions. The structure appears broad and irregular, therefore we model it by using an asymmetric Gaussian. We include in the model the peak at $\sim$31 keV, and the kink at $\sim$46 keV, as Gaussians with resolution fixed to that of the baseline. Given the energy interval of this distribution, we account for the energy dependence of the efficiency in the model. The main structure peaks at 36.27 $\pm$ 0.08 keV and shows an average width of 1.88 $\pm$ 0.11 keV, larger by a factor $\sim$1.4 on the right side , while the weighted average rate among the datasets is 8.3 $\pm$ 0.3 counts/(kg d). 

The monochromatic peak at $\sim$31 keV was observed as well in CUORE-0 demonstrators and past CUORE crystals measurements \cite{LowEnergyTechniques,lowEspectrum} and ascribed to a physics feature. Specifically, in Ref. \cite{lowEspectrum} it was found that a few events were in coincidence with high energy $\gamma$s (507.6 keV and 573.1 keV) associated with $^{121}$Te electron capture (EC), suggesting it to correspond to the binding energy of the K-shell of $^{121}$Sb, consequent to a K-shell EC. Moreover, in Ref. \cite{evidence123te}, a peak at the same energy was used to claim the discovery of $^{123}$Te EC, later refuted in Ref. \cite{123Te2003} and ascribed to the one of $^{121}$Te, activated by neutrons. Its observation in CUORE indicates that the origin of this peak is still present in either TeO$_2$ crystals or the detector setup. We observe its rate to be approximately constant along the data taking with a weighted average among the datasets of 1.35 $\pm$ 0.04 counts/(kg d), while we estimate its mean to be 30.88 $\pm$ 0.09 keV.
The reconstruction of this peak is affected by its proximity to the broad structure around $\sim$36 keV. As a systematic test, we repeat the fit by narrowing the energy range and employing the stricter selection with a 3 keV threshold, to take advantage of the improved resolution. Both tests report a $\sim$30\% lower rate, and a $\sim$0.8\% lower value of the mean, that we account as a systematic uncertainty. 

Focusing on the detectors available down to 3 keV, as in Fig. \ref{fig:LowE3keV}, thanks to a stronger background suppression and improved energy resolution, we observe two excess structures at $\sim$10 and $\sim$13 keV. A similar feature, hidden by the background rise, can be found in the CUORE-0 spectrum in Ref. \cite{LowEnergyTechniques}. We model these features using two Gaussian distributions, fixing the resolution to the one of the baseline, and a linear background. Their means are reconstructed to be (10.3 $\pm$ 0.2) and (12.6 $\pm$ 0.3) keV, with a weighted average rate of (0.66 $\pm$ 0.07) and (0.67 $\pm$ 0.07) counts/(kg d).

We observe a peak on the rising edge close to 3 keV. We modeled it as a Gaussian with resolution fixed to that of the baseline and overlaid it with a linear and an exponential component as the background. We reconstruct a mean and a rate, as the weighted average among the datasets, of 4.69 $\pm$ 0.03 keV and 19.1 $\pm$ 1.0 counts/(kg d), respectively. 

The same peak was observed in past CUORE crystals measurements and the demonstrator Cuoricino. Similar to the peak at $\sim$31 keV, in \cite{lowEspectrum}, a few events in this peak were found to be coincident with higher energy gammas, consistent with the L1-shell EC of $^{121}$Te. However, the observed excess of events in this peak, not coincident with these gammas, was not explicitly attributed to $^{121}$Te EC.

We look for coincidences with the same $^{121}$Te EC $\gamma$ peaks in CUORE data and we find overall 20 events, of which 4 between 30 and 32 keV, 1 between 3 and 6 keV, and the rest spread up to 40 keV. The low significance of this result makes any conclusions drawn from this study unreliable, leaving open the possibility of an alternative explanation.

Finally, concerning the M2 spectrum, we can observe in Fig. \ref{fig:LowE10keV} that after the selection the peaks due to Te X-rays become visible at about 27 and 31 keV. 
The possibility to access this energy regime in CUORE data motivates further investigation towards the reconstruction of a background model, just as the one developed at higher energy \cite{datadriven}. This would pave the way for more new physics searches based on the excess reconstruction over the background \cite{xenonnt}.

\section{Conclusion and Outlook}
In this work we presented the analysis techniques developed to explore the keV-scale energy spectrum of the CUORE experiment acquired over more than 5 years and corresponding to an overall exposure of 2 tonne yr. The adopted methods allowed us to perform a high quality data selection as a function of the energy threshold of interest. We focused on the spectrum down to 10 keV and 3 keV, which are of interest to new physics and rare nuclear decay investigations. 
We studied the performance of the CUORE detectors at both energy thresholds, focusing on the energy resolution, energy calibration, efficiencies, spectrum features and background level. The main information resulting from these selections are summarized in Table \ref{Table:ProcessingResults}. In particular, we highlight the performance improvement, especially in terms of energy resolution, achieved with the stricter selection, in opposition to the loss in exposure.

\begin{table}[htbp]
\centering
\caption{Summary of the main parameters resulting from the selection at low energy of the 2 tonne yr CUORE data release. The reported values correspond to the weighted average among all the datasets, with the propagated uncertainty.}
\begin{tabular}{lccc}
 \hline\hline
\textbf{Parameter} & \textbf{3 keV} & \textbf{10 keV} \\
 \hline
 Number of datasets & 12 &  25 \\ 

 TeO$_2$ exposure [kg yr]& 11.0  &  691  \\ 

 Selected detectors & 1.2\% & 35\%  \\ 

 Baseline FWHM [keV] & 1.18 $\pm$ 0.02 &  2.54 $\pm$ 0.14 \\ 

 Background Level [d.r.u.] & 16 $\pm$ 2  & 2.06 $\pm$ 0.05 \\ 

 Total Efficiency & 0.26 $\pm$ 0.04 & 0.50 $\pm$ 0.02 \\ 
 \hline\hline

\end{tabular}
\label{Table:ProcessingResults}
\end{table}

We studied the main spectral components, which will be fundamental for further investigations about rare nuclear decays (e.g. the aforementioned $^{123}$Te electron capture) and new physics searches, briefly described in the following.

A search for Weakly Interacting Massive Particle (WIMP) dark matter can be performed by looking for the annual modulation signature caused by the seasonal variation in the velocity distribution of WIMP interactions with the detector nuclei due to the Earth's revolution around the sun \cite{Freese_2013}. The WIMP scattering interactions will create observable nuclear recoils, both on Te and O nuclei, below $\sim$50 keV.  Therefore a low threshold is necessary for an increased sensitivity to these scattering interactions \cite{ThesisGPiperno,ThesisSCopello} . CUORE can leverage its 5 years long data taking to observe multiple cycles of this phenomenon. 

Another promising dark matter candidate is the axion, a pseudoscalar boson originally proposed to solve the strong CP problem \cite{axionreview}. This particle can be produced by nuclear reactions in the core of Sun and propagate to the Earth, eventually converting into an electron. The most intense signature is foreseen from the de-excitation of thermally populated first excited state of $^{57}$Fe, which would produce a 14.4 keV monochromatic peak in CUORE crystals. 

Moreover, non-relativistic axions might compose the dark matter halo in the galaxy and interact with matter releasing an amount of energy corresponding to their mass. Performing a scan over accessible energies, CUORE can search for dark matter axions at specific mass values. 

While CUORE already proved excellent performance for monochromatic signatures with the 0$\nu\beta\beta$ search at $\sim$ 2.5 MeV, we can exploit its potential with an analogous signature at the keV-scale, as given by axions interactions \cite{AlbertoRThesis,SamanthaPThesis}. 

More generally, this work enables the development of a detailed background model inspired by Ref. \cite{datadriven} for energies lower than 100 keV, which can be employed for new physics sensitivity studies with future tonne-scale cryogenic calorimeter experiments, such as CUPID (\cite{CUPID-0bkgmodel,CUPID-Mobkgmodel}). 

With this study, we demonstrated that it is possible to operate and study a tonne-scale cryogenic calorimeter experiment over a wide energy range spanning more than 3 orders of magnitude, from the keV-scale, presented in this work, to the MeV-scale, reported in many other works \cite{nature,cuorecollaboration2024nuhuntingseedmatterantimatter,excitedstates,128te,120te,datadriven,fcp}.

Beyond paving the way to new physics investigation at the keV-scale with CUORE data, these results inform planning and goals for future stages of large mass cryogenic calorimeters. 
CUORE will stop data-taking after reaching 3 tonne yr of TeO$_2$ exposure to undergo major upgrades to the cryogenic facility, mainly devoted to the pulse tubes pre-cooling system, that is needed to cool down about a tonne of material down to 4K. This is motivated by the need of its successor, CUPID \cite{CUPIDbaseline,CUPIDbdpt}, for a lower vibration noise to allow the successful operation of light detectors \cite{LDPulseTube,firstmodulecupid,crossprototype}. 
Such an upgrade should be particularly beneficial for low-energy studies in CUORE, serving as a proof of principle for future large-mass cryogenic calorimeters operating in similar facilities. In particular, the noise suppression offered by the cryostat refurbishment can increase the number of available detectors at this energy scale, and thus the sensitivity to beyond Standard Model physics. The background studies can also profit from the improved multi-site event-tagging capability that will be made possible by a higher number of active detectors at lower thresholds. 

The results presented in this work provide key insights for operating the detectors to maximize the performance for keV-scale physics, together with data analysis methods and strategies. The tools we developed will be used to understand the potential for CUORE and future tonne-scale cryogenic calorimeters as multipurpose experiments, demonstrating the scalability of this particle detector's technology in view of both rare decays and dark matter interactions.

\section*{Acknowledgments}
The CUORE Collaboration thanks the directors and staff of the Laboratori Nazionali del Gran Sasso
and the technical staff of our laboratories.
This work was supported by the Istituto Nazionale di Fisica Nucleare (INFN);
the National Science Foundation under Grant Nos. NSF-PHY-0605119, NSF-PHY-0500337,
NSF-PHY-0855314, NSF-PHY-0902171, NSF-PHY-0969852, NSF-PHY-1307204, NSF-PHY-1314881, NSF-PHY-1401832, NSF-PHY-1913374, and NSF-PHY-\\
2412377; Yale University, Johns Hopkins University, and University of Pittsburgh.
This material is also based upon work supported by the US Department of Energy (DOE)
Office of Science under Contract Nos. DE-AC02-05CH11231, and DE-AC52-07NA27344;
by the DOE Office of Science, Office of Nuclear Physics under Contract Nos.
DE-FG02-08ER41551, DE-FG03-00ER41138, DE-SC0012654, DE-SC0020423, DE-SC0019316, and DE-SC0011091.
This research used resources of the National Energy Research Scientific Computing Center (NERSC).
This work makes use of both the DIANA data analysis and APOLLO data acquisition software packages,
which were developed by the CUORICINO, CUORE, LUCIFER, and CUPID-0 Collaborations.
The authors acknowledge the Advanced Research Computing at Virginia Tech and the Yale Center for Research Computing for providing computational resources and technical support that have contributed to the results reported within this paper.

\bibliography{refs.bib}
\end{document}